\documentclass[10pt,twocolumn,showpacs,longbibliography,amsmath,amssymb,osajnl,floatfix,superscriptaddress]{revtex4-1}
\usepackage{graphicx}
\usepackage{dcolumn}
\usepackage{bm}
\usepackage{color}
\usepackage{txfonts}
\usepackage{microtype}
\usepackage{siunitx}
\usepackage{url}
\usepackage{epstopdf}
\usepackage{indentfirst}

\usepackage[colorlinks, linkcolor=black,anchorcolor=black,citecolor=black,filecolor=black,urlcolor=black,bookmarksopen=true]{hyperref}

\usepackage{textcomp}
\usepackage{mathrsfs}
\usepackage{braket}
\usepackage[caption=false,font=normalsize,labelfont=sf,textfont=sf]{subfig}
\usepackage{xcolor} 
\usepackage{amsmath}
\usepackage{upgreek}

\newcommand{\setParDef}{\setlength {\parskip} {0.3pt}}

\begin{document}

\title{Generating Cylindrical Vector $\gamma$ Rays via Beam-Target Interactions: Towards Structured Light at High Energies}

\author{Yue Cao}
\altaffiliation{These authors contributed equally to this work.}
\affiliation{College of Science, National University of Defense Technology, Changsha 410073, China}

\author{Kun Xue}
\altaffiliation{These authors contributed equally to this work.}
\affiliation{Ministry of Education Key Laboratory for Nonequilibrium Synthesis and Modulation of Condensed Matter, State key laboratory of electrical insulation and power equipment, Shaanxi Province Key Laboratory of Quantum Information and Quantum Optoelectronic Devices, School of Physics, Xi'an Jiaotong University, Xi'an 710049, China}

\author{Si-Man Liu}
\affiliation{College of Science, National University of Defense Technology, Changsha 410073, China}

\author{Zhong-Peng Li}	
\affiliation{Ministry of Education Key Laboratory for Nonequilibrium Synthesis and Modulation of Condensed Matter, State key laboratory of electrical insulation and power equipment, Shaanxi Province Key Laboratory of Quantum Information and Quantum Optoelectronic Devices, School of Physics, Xi'an Jiaotong University, Xi'an 710049, China}

\author{Li-Xiang Hu}
\affiliation{College of Science, National University of Defense Technology, Changsha 410073, China}

\author{Xin-Yu Liu}
\affiliation{College of Science, National University of Defense Technology, Changsha 410073, China}

\author{Zhen-Ke Dou}	
\affiliation{Ministry of Education Key Laboratory for Nonequilibrium Synthesis and Modulation of Condensed Matter, State key laboratory of electrical insulation and power equipment, Shaanxi Province Key Laboratory of Quantum Information and Quantum Optoelectronic Devices, School of Physics, Xi'an Jiaotong University, Xi'an 710049, China}

\author{Feng Wan}	
\affiliation{Ministry of Education Key Laboratory for Nonequilibrium Synthesis and Modulation of Condensed Matter, State key laboratory of electrical insulation and power equipment, Shaanxi Province Key Laboratory of Quantum Information and Quantum Optoelectronic Devices, School of Physics, Xi'an Jiaotong University, Xi'an 710049, China}

\author{Qian Zhao}	
\affiliation{Ministry of Education Key Laboratory for Nonequilibrium Synthesis and Modulation of Condensed Matter, State key laboratory of electrical insulation and power equipment, Shaanxi Province Key Laboratory of Quantum Information and Quantum Optoelectronic Devices, School of Physics, Xi'an Jiaotong University, Xi'an 710049, China}

\author{Tong-Pu Yu}
\email{tongpu@nudt.edu.cn}
\affiliation{College of Science, National University of Defense Technology, Changsha 410073, China}

\author{Jian-Xing Li}
\email{jianxing@xjtu.edu.cn}
\affiliation{Ministry of Education Key Laboratory for Nonequilibrium Synthesis and Modulation of Condensed Matter, State key laboratory of electrical insulation and power equipment, Shaanxi Province Key Laboratory of Quantum Information and Quantum Optoelectronic Devices, School of Physics, Xi'an Jiaotong University, Xi'an 710049, China}
\affiliation{Department of Nuclear Physics, China Institute of Atomic Energy, P.O. Box 275(7), Beijing 102413, China}

\date{\today}

\begin{abstract}

Structured $\gamma$ rays, particularly cylindrical vector $\gamma$ rays, offer promising tools for sub-nuclear imaging and polarization-sensitive probes in fundamental research and applications, but conventional optical methods face great challenges at such photon energy. Here, we put forward a novel method generating such $\gamma$ rays through relativistic beam-target interactions. For instance, radially polarized $\gamma$ rays can be generated by using a dense electron beam striking a multifoil target. We find that the radial polarization is transferred from the generated coherent transition radiation (CTR) fields to  $\gamma$ photons through nonlinear Compton scattering, with the high polarization preserved by phase matching. Three-dimensional spin-resolved simulations demonstrate radial polarization degrees approaching 60\%. Furthermore, these $\gamma$ rays can decay into azimuthally spin-polarized positrons via the nonlinear Breit-Wheeler process, with their spins aligning along the CTR magnetic field. Our work extends the concept of structured light into the $\gamma$-ray regime, offering new prospects for broad fields such as nuclear structure probing, fundamental symmetries tests, polarization-sensitive studies in extreme conditions, and laboratory astrophysical observations.

\end{abstract}

\maketitle

\setParDef
Structured light, characterized by spatiotemporally modulated distributions of intensity, phase, and polarization, has become a powerful tool, driving both fundamental discoveries (e.g., spin-orbit interaction phenomena~\cite{bliokh2015spin}) and technological applications (e.g., super-resolution lithography~\cite{xie2014harnessing}) across multiple fields~\cite{Rubinsztein2017Roadmap, forbes2021structured, Marco2025Trends}. A prominent example is cylindrical vector beams (CVBs), which exhibit cylindrical symmetry in their polarization topology, encompassing radially, azimuthally, or hybridly polarized beams. Within the optical regime ($\sim$ eV-level), CVBs have enabled breakthroughs in areas like optical trapping and tweezing~\cite{Yang2021Optical, Skelton2013Trapping, Huang2012Optical, Donato2012Optical, Moradi2019Efficient}, as well as confocal microscopy~\cite{Liu2022Super}, pushing spatial resolution beyond the conventional diffraction limit. However, the optical wavelength scale (hundreds of nanometers) makes it challenging to probe or manipulate matter at sub-nanometer dimensions. Extending CVBs from the optical to $\gamma$-ray regime provides access to extremely short wavelengths, enabling spatial resolutions sufficient to probe or manipulate nuclear and subnuclear structures, such as the spin and parity of nuclei~\cite{Iliadis2021Linear, Zilges22PPNP} and the internal spin structure of the proton~\cite{Thiel2012Well, Gottschall2014First}. Remarkably, recent astronomical observations from supernova remnants have reported the natural emission of high-energy (over kiloelectronvolt) photons with radial and azimuthal polarization, offering unprecedented insights into cosmic magnetic field configurations and the dynamics of relativistic jets~\cite{Prokhorov2024Evidence}. However, the absence of laboratory-based high-energy-CVB sources poses a critical bottleneck in experimentally validating these astrophysical phenomena, highlighting the imperative need for controlled generation techniques.

Current optical methods for generating CVBs can be categorized as active or passive, depending on whether amplifying media are employed~\cite{zhan2009cylindrical,chen2018vectorial,wang2021generation}. Active methods generate CVBs directly within the laser cavity through controlled mode selection and phase modulation, enabled by tailored optical components such as intracavity axial birefringent or dichroism crystals~\cite{yonezawa2006generation,machavariani2007birefringence,bisson2006radially}, q-plates~\cite{sanchez2015performance,naidoo2016controlled}, and reconfigurable multicore fibers~\cite{sun2012low,lin2015tungsten,mao2021generation}. In contrast, passive methods modulate homogeneously polarized input beams through external optical devices, including phase-only spatial light modulators~\cite{neil2002method, maurer2007tailoring, ruiz2013highly,han2013vectorial}, digital micro-mirror devices~\cite{mitchell2016high}, custom-designed fiber gratings~\cite{hirayama2006generation,mao2021generation}, and metasurfaces~\cite{liu2014realization, yue2016vector}. 
However, these methods face challenges in the $\gamma$-ray regime: with wavelengths far smaller than atomic dimensions, $\gamma$ rays exceed the modulation capabilities of conventional materials based on macroscopic optical mechanisms such as interference and refraction~\cite{chen2018vectorial,zhao2025research}.

Meanwhile, homogeneously polarized $\gamma$ rays can be generated via several mechanisms, including bremsstrahlung~\cite{olsen1959photon, Kuraev2010Bremsstrahlung}, linear Compton scattering (LCS)~\cite{Howell2021International}, and nonlinear Compton scattering (NCS)~\cite{li2020polarized, Xue2020Generation, wang2024manipulation}. 
For instance, bremsstrahlung can produce either circular or linear polarization. The former arises from longitudinally spin-polarized electrons interacting with metallic targets via incoherent bremsstrahlung~\cite{Giulietti2008Intense, albert2016applications}, while the latter is achieved via coherent bremsstrahlung in crystals, where the periodic lattice enables phase-matched emission~\cite{Uggerhoj2005The, ter1972high, Lohmann1994Linearly}. In the LCS regime, the polarization (linear/circular) of the produced $\gamma$ rays is inherited from the incident laser pulse interacting with a relativistic electron beam~\cite{baier1973radiation, ritus1985quantum, sokolov1986radiation, khokonov2010length, Omori2006efficient, Alexander2008Observation, Petrillo2015Polarization}. In the NCS regime, circularly polarized $\gamma$ rays are produced by collisions between longitudinally spin-polarized electron beams and ultraintense laser pulses~\cite{li2020polarized}, whereas linear polarization does not require polarized electron beams~\cite{Wan2020High}. 
However, these approaches generally face difficulties in generating $\gamma$ rays with spatially structured polarization. 
Notably, recent theoretical work has demonstrated that terahertz waves can manipulate the spatial spin structure of relativistic polarized lepton beams~\cite{li2025generation, li2025ultrafast}.
However, this method has not yet been extended to photons. 
Therefore, the efficient generation of $\gamma$ rays with controlled spatial polarization, particularly cylindrical vector modes, remains a significant challenge.

In this Letter, we investigate the generation of cylindrical vector $\gamma$ rays through relativistic charged particle beam-target interactions, which provide a distinctive route to generating the spatially structured fields required for polarization control. For instance, radially polarized $\gamma$ rays can be generated by using a dense electron beam striking a multifoil target; see Fig.~\ref{fig1}(a). (An alternative beam-cone configuration is presented in the Supplemental Material (SM)~\cite{SM}.) The radial polarization is transferred from the generated coherent transition radiation (CTR) fields to $\gamma$ photons through NCS [see Fig.~\ref{fig1}(b)], with phase matching preserving high polarization by suppressing the polarization cancellation. Additionally, by using a rotating electron beam, azimuthal polarization components can be introduced to the $\gamma$-ray polarization through the transfer of azimuthal momentum. 
Three-dimensional spin-resolved quantum electrodynamics (QED) particle-in-cell (PIC) simulations demonstrate the generation of high-energy $\gamma$ rays with a high radial polarization degree approaching 60\%. 
Furthermore, these $\gamma$ rays can subsequently produce azimuthally spin-polarized positrons via the nonlinear Breit-Wheeler (NBW) process, with their spins aligning along the azimuthal CTR magnetic field; see Fig.~\ref{fig1}(c). Such positron beams may find applications in areas including polarized deep inelastic scattering~\cite{Anselmino1995The, Hughes1999Spin, Bluemlein2013The, Abelev2009Longitudinal, Adamczyk2014Measurement} and chiral-selective chemistry~\cite{Rosenberg2011Spin, Bonner1991The, kessler85polarized}. 
Our study contributes to extending the concept of CVBs to high-energy $\gamma$ rays and relativistic lepton regimes, potentially enabling new applications in broad fields such as high-energy physics~\cite{moortgat2008polarized, Akbar2017Measurement, bragin2017high}, nuclear physics~\cite{Iliadis2021Linear, Zilges22PPNP, Thiel2012Well, Gottschall2014First}, laboratory astrophysics~\cite{Prokhorov2024Evidence}, etc.

\begin{figure}	[t]	 
	\setlength{\abovecaptionskip}{-0.2cm}
	\setlength{\belowcaptionskip}{-0.3cm}
	\centering
	\includegraphics[width=1.0\linewidth]{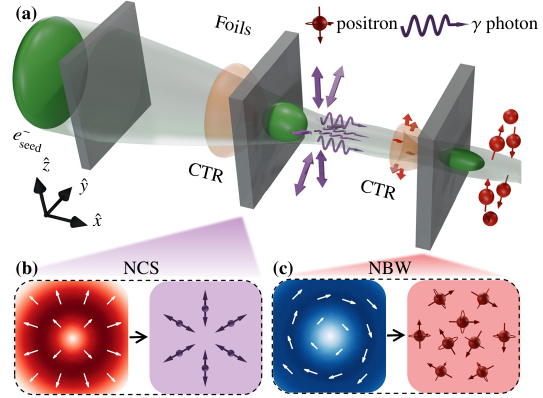}
	\begin{picture}(300,0)	
	\end{picture}
	\caption{Schematic illustration. (a) An initially unpolarized seed electron beam $e^-_{\rm seed} $ (green) propagating along the $+x$ direction interacts sequentially with multiple foils, leading to the emission of polarized $\gamma$ photons via NCS and the subsequent generation of polarized positrons via the NBW process. The light green profile illustrates the beam focusing, while the arrows on the positrons indicate their spin polarization. For simplicity, only three foils are shown. 
	(b) In the NCS process, the CTR electric field, whose magnitude and direction are shown by the red heatmap and white arrows, determines the polarization of the emitted photons (purple double-headed arrows). 
	(c) In the NBW process, the CTR magnetic field, indicated by the blue heatmap and white arrows, governs the spin of the generated positrons (red arrows). 
	}
\label{fig1}
\end{figure}

 \textit{Simulation methods  and setup \textemdash} We simulate the beam-plasma interactions using spin-resolved QED-PIC code SLIPs~\cite{wan2023simulations}, which incorporates Monte Carlo algorithms for NCS and NBW processes.
These two processes are characterized by the nonlinear QED parameters $\chi_e\equiv|e|\hbar/(m_e^3c^4)\sqrt{-(F_{\mu\nu}p^{\nu})^2}$ and $\chi_\gamma\equiv|e|\hbar/(m_e^3c^4)\sqrt{-(F_{\mu\nu}k^{\nu})^2}$, respectively~\cite{baier1998Electromagnetic}. Here, $F_{\mu\nu}$ is the electromagnetic field tensor, $p^\nu$ and $k^\nu$ are the four-momentum of electrons and $\gamma$ photons, respectively, $\hbar$ is the reduced Planck constant, $e$ and $m_e$ are the electron charge and mass, respectively, and $c$ is the speed of light in vacuum. 
When a photon is emitted, its polarization is described by the Stokes parameters ($\xi_1$, $\xi_2$, $\xi_3$), defined relative to the instantaneous frame ($\hat{\mathbf{k}}_\gamma$, $\hat{\mathbf{e}}_1$, $\hat{\mathbf{e}}_2$), with $\hat{{\mathbf{e}}}_1=\hat{\mathbf a}-\hat{\mathbf v}(\hat{\mathbf v}\cdot\hat{\mathbf a})$ and $\hat{{\mathbf e}}_2=\hat{\mathbf v}\times\hat{\mathbf a}$. Here, $\hat{\mathbf{a}}$, $\hat{\mathbf{v}}$, and $\hat{\mathbf{k}}_\gamma$ are unit vectors along the electron acceleration, velocity, and photon wave vector, respectively. To evaluate the angular distribution of polarization, the Stokes parameters of individual photons with the same propagation direction are first transformed from their instantaneous frames ($\hat{\mathbf{k}}_\gamma$, $\hat{\mathbf{e}}_1$, $\hat{\mathbf{e}}_2$) to a common observation frame ($\hat{\mathbf{k}}_\gamma$, $\hat{\mathbf{o}}_1$, $\hat{\mathbf{o}}_2$), and then averaged. The degree of polarization at each $\hat{\mathbf{k}}_\gamma$ is given by $P_r \equiv \sqrt{\overline{\xi_1}^2 + \overline{\xi_3}^2}$. Field ionization is modeled using a hybrid approach including tunnel ionization~\cite{Ammosov1986tunnel} and barrier suppression ionization~\cite{posthumus1997molecular}. 

The size of the simulation box is $x\times y \times z=6.6~\mathrm{\upmu m} \times 12~\mathrm{\upmu m} \times 12~\mathrm{\upmu m}$, with spatial resolution $\Delta x \times \Delta y \times \Delta z = 0.01~\mathrm{\upmu m} \times 0.02~\mathrm{\upmu m} \times 0.02~\mathrm{\upmu m}$. 
A moving window is employed, starting at $t = 6T_0$ and propagating along $+\hat{x}$, where $T_0= 1$ $\mathrm{ \upmu m}/c$ represents the time for light to travel 1 $\rm \upmu m$. 
A sequence of carbon foils is positioned perpendicular to the $x$-axis, starting at $x = 5\, {\rm \upmu m}$, each with thickness $d = 0.5~\mu\mathrm{m}$, interfoil distance $l = 5.0~\mu\mathrm{m}$, and initial density $n_t = 4\times10^{28}\, \mathrm{m}^{-3}$. A seed electron beam with $570\ \rm pC$ charge, $1.5\ \rm GeV$ energy, 5\% energy spread, and 0.5$^{\circ}$ angle spread propagates along $+\hat{{x}}$.  Its spatial distribution follows a Gaussian distribution $n_\mathrm{seed} = n_\mathrm{b0}\exp\left(-{x^2}/{2\sigma_\parallel^2}-{r^2}/{2\sigma_\perp^2}\right)$, where $n_{\rm b0}$ is the maximum density of the seed beam, $r=\sqrt{y^2+z^2}$, $\sigma_\parallel$ = 0.55~$\upmu$m, and $\sigma_\perp$ = 1.0~$\upmu$m. 
Comparable beam parameters are achievable with advanced conventional or laser-driven accelerators~\cite{yakimenko2019FACET,  yakimenko2019prospect, Emma2025Experimental, Clarke2022advanced, babjak2024direct, Pukhov1999Particle, aniculaesei2023acceleration}. For instance, the FACET-II facility can generate 10 GeV, 3 nC electron beams~\cite{Emma2025Experimental}, with a planned upgrade to 5 nC~\cite{yakimenko2019FACET}. Such beams can be compressed to 0.5 $\upmu$m longitudinally and 5 $\upmu$m transversely~\cite{yakimenko2019FACET}, 
and further compressed transversely to $<1\, \mathrm{\upmu m}$ using techniques like plasma lenses~\cite{doss2019laser} and magnetic pinching~\cite{zhu2023magnetic}.  
Notably, low-$Z$ target materials are preferable for suppressing bremsstrahlung and Bethe-Heitler (BH) processes, since both cross-sections scale approximately as $Z^2$~\cite{heitler1954quantum,SM}. In the simulations, each cell contains 8 macroparticles for beam electrons and 4 for carbon atoms.

\begin{figure}	[t]	 
	\setlength{\abovecaptionskip}{-0.2cm}
	\setlength{\belowcaptionskip}{-0.3cm}
	\centering
	\includegraphics[width=1.0\linewidth]{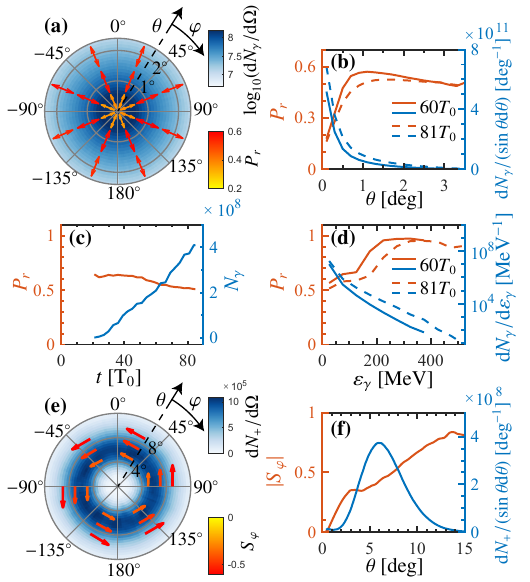}

	\begin{picture}(300,0)	
	\end{picture}
	\caption{(a) Angle-resolved distribution $\log_{10}\left(\rm{d}\it N_\gamma/\rm{d}\Omega \right)$ (blue heat map) and average polarization $P_r$ (red double-headed arrows) of all emitted $\gamma$ photons at $t=60T_0$ vs the polar angle $\theta$ and the azimuth angle $\varphi$, respectively. Here, the direction of the red double-headed arrows represents the polarization direction of the $\gamma$ rays; ${\rm d} \Omega$ = $\sin\theta{\rm d}\theta{\rm d}\varphi$, $\theta$ = $0^{\circ}$ in $+\hat{{x}}$, and $\varphi$ = $0^{\circ}$ in $+\hat{{y}}$. 
	(b) Angle-resolved polarization degree $P_r$ (red) and distribution $\rm{d}\it N_\gamma/(\sin\theta\rm{d}\theta)$ (blue) of all emitted $\gamma$ photons vs $\theta$ at $t = 60T_0$ (solid) and $81T_0$ (dotted), respectively. 
	(c) Temporal evolution of the number (blue) and the polarization degree $P_r$ (red) for $\gamma$ photons within $1^\circ<\theta<1.5^\circ$. 
	(d) Energy-resolved polarization degree $P_r$ (red) and distribution ${\rm d}N_\gamma/{\rm d}\varepsilon_\gamma$ (blue) of $\gamma$ photons within $1^\circ<\theta<1.5^\circ$ vs the photon energy $\varepsilon_\gamma$ at $t = 60T_0$ (solid) and $81T_0$ (dotted), respectively. 
	(e) Angle-resolved positron distribution (blue heat map) and azimuthal spin-polarization $S_{\varphi}$ (red arrows) vs $\theta$ and $\varphi$. Here, $S_\varphi \equiv \mathbf{S} \cdot \hat{e}_\varphi$ with $\hat{e}_\varphi=(0,-p_z/p_\perp, p_y/p_\perp)$ and $p_\perp=\sqrt{p_y^2+p_z^2}$.
	(f) Angle-resolved positron distribution ${\rm d}N_+/(\sin\theta{\rm d}\theta)$ (blue) and azimuthal polarization $S_{\varphi}$ (red) vs $\theta$.
	}
	\label{fig2}
\end{figure}

\begin{figure}	[t]	 
	\setlength{\abovecaptionskip}{-0.4cm}
	\setlength{\belowcaptionskip}{-0.3cm}
	\centering
	\includegraphics[width=1.0\linewidth]{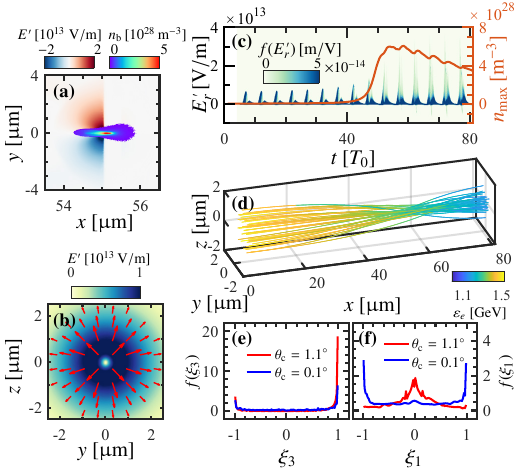}
	\begin{picture}(300,0)	
	\end{picture}
	\caption{Spatial distribution of the effective field $E^{\prime}$ in (a) the $xy$ plane ($z = 0$~$\upmu$m) and (b) the $yz$ plane ($x \approx 55\, {\rm \upmu m}$) at $t=57T_0$. 
Panel (a) includes the seed electron beam density, while red arrows in (b) indicate the direction of $E'$.
(c) Temporal evolution of the normalized electron distribution vs the radial component of the effective field $E^{\prime}_r$ (heatmap), with the distribution function $f(E^{\prime}_r)={\rm d}N_e/(N_0{\rm d}E^{\prime}_r)$ and the total number of sampled electrons $N_0$. The red solid line shows the evolution of the maximum beam density $n_{\rm max}$. (d) Sampled electron trajectories color-coded by energy $\varepsilon_e$. Normalized electron distributions as functions of the Stokes parameters, $ f(\xi_1) = {\rm d}N_{\rm e}/N_0 {\rm d}\xi_1$ (e) and $ f(\xi_3) = {\rm d}N_{\rm e}/N_0 {\rm d}\xi_3$ (f), for photons within the angular range of ($ |\theta - \theta_{\rm c}| < 0.1^{\circ} $, $ |\varphi| < 0.5^{\circ} $), centered at $ \theta_{\rm c} = 0.1^{\circ} $ (blue) and $1.1^{\circ} $ (red), respectively.
Here, $\xi_3 = \pm 1$ indicates that the direction of polarization is parallel ($+1$) or perpendicular ($-1$) to  $\hat{\mathbf{e}}_1$, while $\xi_1 = \pm 1$ corresponds to polarization along directions oriented at $\pm 45^\circ$ with respect to $\hat{\mathbf{e}}_1$.
}
\label{fig3}
\end{figure}

\textit{Properties of radially polarized $\gamma$ rays and resulting positrons  \textemdash} The emitted $\gamma$ rays exhibit radial polarization and a cylindrically symmetric angular distribution; see Fig.~\ref{fig2}(a). 
Given this symmetry, we will focus on the polar angle ($\theta$) dependence. As $\theta$ increases from 0$^\circ$, the photon yield gradually decreases; see Fig.~\ref{fig2}(b). The polarization degree reaches a maximum of about 57\% at $\theta \approx 1.1^\circ$, and then gradually decreases at larger polar angles. A sharp drop in polarization near $\theta \approx 0^\circ$ arises from polarization cancellation, where photons along this axis represent a superposition of different polarization states; see physical reasons in Figs.~\ref{fig3}(e) and (f). This polarization distribution closely resembles the V-point polarization singularity observed in optical CVBs~\cite{milione2011higher}. 

As shown in Fig.~\ref{fig2}(c), with $\theta \in (1^\circ, \, 1.5^\circ)$, the photon number increases over time due to cumulative emission, while the average polarization degree slightly decreases, as later-emitted photons exhibit lower polarization. These photons exhibit an exponential energy spectrum with a cutoff near 400 MeV; see Fig.~\ref{fig2}(d). The polarization degree increases from 60\% at the MeV range to nearly 100\% around 200 MeV, consistent with the theoretical prediction that $P_r$ grows with $\varepsilon_\gamma / \varepsilon_e$ increasing from zero~\cite{Xue2020Generation}. Furthermore, the photons at $t = 81T_0$ exhibit a higher cutoff energy and greater flux than those at $t = 60T_0$, albeit with a reduced polarization degree, highlighting the need to optimize the number of foils according to specific application demands; see Figs.~\ref{fig2}(b) and (d). 
These parameters fall within the ranges required to probe nuclear and subnuclear structure: energies of hundreds of MeV to several GeV and a polarization of  $\sim 60\%$ ~\cite{Iliadis2021Linear, Zilges22PPNP, Gottschall2014First, Thiel2012Well}. Furthermore, their radial polarization provides an additional degree of freedom: the spatial distribution of the polarization, which may influence experimental observations like the angular distribution of scattered particles and could yield signatures of new physics. 

At $t = 60T_0$, the photons within $1^\circ < \theta < 1.5^\circ$ have a root-mean-square angular divergence of $15.4\times15.4$ $\rm mrad^2$, and spatial dimensions of $0.27$ $\rm \upmu m$ (longitudinal) and $0.36\, {\rm \upmu m} \times 0.36\, {\rm \upmu m}$ (transverse). The brilliances at $\varepsilon_\gamma = 50$, 100, 150, and 200 $\rm MeV$ are $1.9 \times 10^{26}$, $3.2 \times 10^{25}$, $7.9 \times 10^{24}$, and $3.3 \times 10^{24}\ \text{photons}/(\rm s\ mm^2 mrad^2 \times0.1\%\ bandwidth)$, respectively. These values significantly exceed those typically achieved in laser-electron beam collision schemes by over five orders of magnitude~\cite{li2020polarized, Xue2020Generation}, owing to the ultrastrong electromagnetic fields and improved beam collimation from the self-focusing effect in the beam-target interaction~\cite{sampath2021extremely, zhu2023efficient, Cui2025Generation}.

The generated photons can further decay into electron-positron pairs via the NBW process in the CTR field, which becomes significant when $\chi_\gamma \gtrsim 1$~\cite{piazza2012extremely}. As an illustrative case for discussing the positron generation and polarization, Figs.~\ref{fig2}(e) and (f) present results for a 5-nC, 10-GeV seed electron beam, with all other parameters identical to those in Figs.~\ref{fig2}(a)-(d). The generated positrons exhibit cylindrical symmetry in both angular and polarization distributions, with their spin polarization aligned antiparallel to the azimuthal direction; see Fig.~\ref{fig2}(e). The total positron charge reaches approximately  38~pC. The polar angle peaks at $\theta = 6.0^\circ$ with a full width at half maximum (FWHM) of about $5.1^\circ$, and the degree of positron polarization  $|S_\varphi|$ increases from zero to $\sim 80\%$ as $\theta$ increases; see Figs.~\ref{fig2}(f). Moreover, the seed electron beam undergoes spontaneous radiative polarization, with the low-energy fraction ($\sim 10\%$ of the beam) exhibiting a polarization of $43\%$ ~\cite{SM}.

\textit{Polarization Mechanisms of $\gamma$ rays and Positrons \textemdash} 
As the seed electron beam traverses the multiple foils, it ionizes the targets and excites intense near-field CTR at each surface due to the abrupt change in dielectric constant~\cite{jackson2021classical, sampath2021extremely}. The CTR field exhibits radial polarization and takes the form $\mathbf{E}_\mathrm{CTR} = E_r\hat{\mathbf{e}}_r + E_x\hat{\mathbf{e}}_x$ and $\mathbf{B}_\mathrm{CTR} = -B_\vartheta\hat{\mathbf{e}}_\vartheta$ in a cylindrical coordinate frame aligned with the $x$-axis, where the amplitudes $E_r$, $E_x$, and $B_\vartheta$ are all proportional to the beam density $n_{b0}$~\cite{sampath2021extremely}. For relativistic electrons propagating along the $x$-axis, the effective electric field can be written as $\mathbf{E}^{\prime} = \mathbf{E}_\perp+\mathbf{v}\times\mathbf{B}\approx(E_r + c B_\vartheta) \hat{\mathbf{e}}_r$, where $\mathbf{E}_\perp$ is the transverse electric field. Due to phase matching between the electron beam and the CTR field, the beam consistently experiences outward radial effective field; see Figs.~\ref{fig3}(a) and (b). Therefore, the electrons move radially inward and become focused, reaching the maximum density at $t\approx60T_0$; see Figs.~\ref{fig3}(c) and (d). Note that the rising electron density amplifies the CTR field, which in turn further promotes electron density growth. Our simulations show that the effective field can reach up to $3 \times 10^{13}$ V/m.

In such a strong field, $\gamma$-ray emission becomes significant via the NCS process~\cite{mackenroth2013nonlinear, li2015attosecond, yu2024bright, zhao2019ultra}, where the nonlinear QED parameter reaches $\chi_e = \gamma_e|\mathbf{E}'|/E_s \sim 0.1$. Here, $\gamma_e$ is the Lorentz factor of electrons, and $E_s=m_e^2c^3/e\hbar \approx 1.3\times10^{18}$~V/m is the QED critical field strength~\cite{schwinger1951gauge}. To analyze the polarization, we first define an appropriate observation frame. For electrons initially moving along the $x$-axis (i.e., the angle spread of the beam $\Delta\theta_0=0$), their trajectories are confined to the local $\hat{{\mathbf{e}}}_x$-$\hat{{\mathbf{e}}}_r$ plane, owing to the cylindrical symmetry of the CTR field. When a $\gamma$ photon is emitted, assuming that the parent electron has velocity $\hat{\mathbf{v}}=\cos\theta \hat{\mathbf{e}}_x+\sin\theta \hat{\mathbf{e}}_r$ and the acceleration $\hat{\mathbf a} \approx \mathbf{E}'/|\mathbf{E}'| \approx -\hat{\mathbf{e}}_r$, the instantaneous frame is given by $\hat{\mathbf{k}}_\gamma\approx\hat{\mathbf v}$, $\hat{{\mathbf{e}}}_1^0 =\hat{\mathbf a}-\hat{\mathbf v}(\hat{\mathbf v}\cdot\hat{\mathbf a}) = \cos^2\theta\hat{{\mathbf e}}_r+\sin\theta\cos\theta\hat{{\mathbf e}}_x \approx \hat{{\mathbf e}}_r \parallel \mathbf{E}'$, and $\hat{{\mathbf{e}}}_2^0 = \hat{\mathbf v} \times \hat{\mathbf a} = -\cos\theta\hat{{\mathbf e}}_x\times\hat{{\mathbf e}}_r \approx \hat{{\mathbf{e}}}_\vartheta$, where the first approximation follows from the small emission angle ($\sim 1/\gamma_e \ll 1$)~\cite{jackson2021classical}, and the latter two are supported by simulation results with $\theta<3^\circ$; see Fig.~\ref{fig2}(c). Since photons with the same wavevector $\hat{\mathbf{k}}$ share a common instantaneous frame ($\hat{\mathbf{k}}_\gamma$, $\hat{\mathbf{e}}_1^0$, $\hat{\mathbf{e}}_2^0$), we adopt this frame as the observation frame ($\hat{\mathbf{k}}_\gamma$, $\hat{\mathbf{o}}_1$, $\hat{\mathbf{o}}_2$). According to polarization-resolved NCS theory~\cite{li2020polarized,Xue2020Generation,SM}, unpolarized electrons undergoing planar motion yield average Stokes parameters $\overline{\xi}_1 = \overline{\xi}_2=0$ and $\overline{\xi}_3>0$, indicating radial polarization of the emitted $\gamma$ rays along $\hat{\mathbf{e}}_1^0 \approx \hat{\mathbf{e}}_r \parallel \mathbf{E}'$. For beams with a small angular divergence $\Delta\theta_0$, this radial polarization behavior persists, but the degree of polarization decreases for the photons with $\theta \lesssim \Delta\theta_0$; see Fig.~\ref{fig2}(b). This decrease arises from the fact that photons with the same $\hat{\mathbf{k}}_\gamma$ can originate from parent electrons moving in different local $\hat{\mathbf{e}}_x$-$\hat{\mathbf{e}}_r$ planes, resulting in polarization vectors along different $\hat{\mathbf{e}}_1$ directions that partially cancel each other; see the numerical results in Figs.~\ref{fig3}(e) and (f). Fortunately, phase matching enables continuous focusing, allowing electrons to accumulate substantial radial momentum. The resulting large polar angle $\theta$ of the emitted photons suppresses the polarization cancellation, enabling a high polarization. By contrast, the direct collision of an electron beam with a radially polarized laser is ineffective for generating such $\gamma$ rays (see the SM~\cite{SM}). 

\begin{figure}	[t]	 
	\setlength{\abovecaptionskip}{-0.4cm}
	\setlength{\belowcaptionskip}{-0.4cm}
	\centering
	\includegraphics[width=0.950\linewidth]{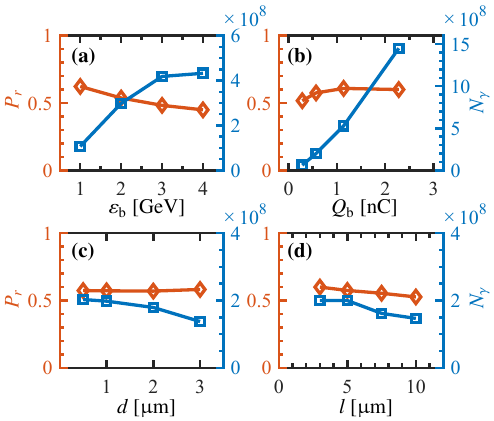}
		\begin{picture}(300,0)
		\end{picture}
	\caption{Effects of (a) beam energy $\varepsilon_{\rm b}$, (b) beam charge $Q_{\rm b}$, (c) foil thickness $d$, and (d) interfoil distance $l$ on the polarization $P_r$ and the number $N_\gamma$ for photons within $1^{\circ}<\theta<1.5^{\circ}$. All data are taken at the time when the electron beam reaches its peak density.
	}
	\label{fig4}
\end{figure}

The emitted $\gamma$ photons can further decay into electron-positron pairs via the NBW process~\cite{ridgers2012dense, piazza2016nonlinear, xie2017electron, vranic2018mult, zhu2016dense}. Given the cylindrical symmetry of both the CTR field and the angular distribution of the emitted $\gamma$ photons, the generated positron spins tend to align along $-\hat{\mathbf{b}}_{+} \equiv \hat{\mathbf{v}}_+\times\hat{{\mathbf a}}_+/|\hat{\mathbf{v}}_+\times\hat{{\mathbf a}}_+| \parallel -\hat{\mathbf{k}}\times \hat{\mathbf{a}}_+ \parallel -\hat{\mathbf{v}} \times \mathrm{E}' = -(\cos\theta \hat{\mathbf{e}}_x+ \sin\theta \hat{\mathbf{e}}_r) \times (E_r-cB_\vartheta)\hat{{\mathbf e}}_r \parallel \hat{{\mathbf e}}_\vartheta \parallel \mathbf{B}_\mathrm{CTR}$~\cite{SM}, where $\hat{\mathbf{v}}_+$ and $\hat{{\mathbf a}}_+$ are the unit vectors along the positron velocity and acceleration, respectively. Owing to the phase matching between the $\gamma$ photons and the CTR field, the photons experience a nearly fixed $\mathbf{B}_\mathrm{CTR}$ direction, enabling the generation of spin-polarized positrons along the azimuthal direction. 
 
\textit{Experimental Feasibility Analysis \textemdash} To validate the experimental feasibility of our scheme, we investigate the influence of key parameters on the photon number $N_\gamma$ and polarization $P_r$, as summarized in Fig.~\ref{fig4}. Increasing the seed beam energy $\varepsilon_{\rm b}$ increases the photon number $N_\gamma$ while slightly reducing the polarization $P_r$ [Fig.~\ref{fig4}(a)], as a higher $\varepsilon_{\rm b}$ increases $\chi_e$, enhancing the radiation probability while lowering polarization~\cite{Xue2020Generation, SM}. Conversely, increasing the beam charge $Q_{\rm b}$ enhances $N_\gamma$ almost linearly and slightly raises $P_r$ [Fig.~\ref{fig4}(b)]. This is because a higher $Q_{\rm b}$ generates stronger CTR fields, which not only boost $N_\gamma$ by increasing $\chi_e$ but also improve the beam focusing. This enhanced focusing, in turn, more effectively suppresses the polarization cancellation, leading to a higher net polarization. Increasing the foil thickness $d$ slightly decreases the photon number $N_\gamma$ while leaving the polarization $P_r$ nearly unchanged; see Fig.~\ref{fig4}(c). This results from the reduced interfoil gap ($l - d$), which shortens the interaction length between the beam and the interfoil field. As the interfoil distance $l$ increases, the cumulative beam focusing weakens because the beam spends more time between foils without the strong focusing fields. Consequently, both $N_\gamma$ and $P_r$ decrease slightly due to the reduced photon yield and diminished suppression of the polarization cancellation. Moreover, we investigated the influence of the seed beam energy spread and the target density, finding their effects negligible since both have little impact on the generated CTR field and radiation probability~\cite{SM}.

We have also investigated several alternative configurations, further confirming the promising potential of our approach for generating structured $\gamma$ rays (see SM~\cite{SM}, Secs.~VI--VIII). 
For instance, beam-cone interactions offer an alternative approach for generating radial polarization, which arises from the azimuthal magnetic field induced by return currents in the cone walls. 
Additionally, rotating electron beams can be used to generate hybrid CVB modes, as photons inherit azimuthal momentum from the electrons, leading to observable azimuthal polarization components in the angle-resolved distribution. 
In principle, our method is also applicable to other charged particle beams like dense positron beams.
Furthermore, the structure of the self-generated fields can be tailored by modifying the target design, opening up the possibility of generating other forms of structured light in the $\gamma$-ray regime, such as that with customized intensity distributions. 

In conclusion, we put forward a novel method based on relativistic beam-target interactions for generating $\gamma$ rays and positron beams with structured polarization, which remains challenging to achieve with existing optical or laser-based methods. Our work extends the concept of structured light into the high-energy photon regime, opening new opportunities for broad fields such as nuclear structure probing, fundamental symmetries tests, polarization-sensitive studies in extreme conditions, and laboratory astrophysical observations.

\hspace*{\fill}

\textit{Acknowledgements\textemdash} This work is supported by the National Natural Science Foundation of China (Grants Nos. 12135009, 12275209, 12375244, 12425510, 12441506, 12447106, 12475249, and U2267204), the National Key Research and Development (R\&D) Program (Grant Nos. 2024YFA1610900, 2024YFA1612700), the Science Challenge Project (No. TZ2025012), the Innovative Scientific Program of CNNC, the Natural Science Foundation of Hunan Province of China
(Grant No. 2025JJ30002), the China Postdoctoral Science Foundation (Grant No. 2024M762568) and the Fundamental Research Funds for the Central Universities of Ministry of Education of China (Grant Nos. xzy012025079, xzy012023046). Yue Cao thanks Prof. Jian-Xing Li for his hospitality at Xi'an Jiaotong University.

\bibliography{mybib}

\begin{thebibliography}{99}%
\makeatletter
\providecommand \@ifxundefined [1]{%
 \@ifx{#1\undefined}
}%
\providecommand \@ifnum [1]{%
 \ifnum #1\expandafter \@firstoftwo
 \else \expandafter \@secondoftwo
 \fi
}%
\providecommand \@ifx [1]{%
 \ifx #1\expandafter \@firstoftwo
 \else \expandafter \@secondoftwo
 \fi
}%
\providecommand \natexlab [1]{#1}%
\providecommand \enquote  [1]{``#1''}%
\providecommand \bibnamefont  [1]{#1}%
\providecommand \bibfnamefont [1]{#1}%
\providecommand \citenamefont [1]{#1}%
\providecommand \href@noop [0]{\@secondoftwo}%
\providecommand \href [0]{\begingroup \@sanitize@url \@href}%
\providecommand \@href[1]{\@@startlink{#1}\@@href}%
\providecommand \@@href[1]{\endgroup#1\@@endlink}%
\providecommand \@sanitize@url [0]{\catcode `\\12\catcode `\$12\catcode
  `\&12\catcode `\#12\catcode `\^12\catcode `\_12\catcode `\%12\relax}%
\providecommand \@@startlink[1]{}%
\providecommand \@@endlink[0]{}%
\providecommand \url  [0]{\begingroup\@sanitize@url \@url }%
\providecommand \@url [1]{\endgroup\@href {#1}{\urlprefix }}%
\providecommand \urlprefix  [0]{URL }%
\providecommand \Eprint [0]{\href }%
\providecommand \doibase [0]{http://dx.doi.org/}%
\providecommand \selectlanguage [0]{\@gobble}%
\providecommand \bibinfo  [0]{\@secondoftwo}%
\providecommand \bibfield  [0]{\@secondoftwo}%
\providecommand \translation [1]{[#1]}%
\providecommand \BibitemOpen [0]{}%
\providecommand \bibitemStop [0]{}%
\providecommand \bibitemNoStop [0]{.\EOS\space}%
\providecommand \EOS [0]{\spacefactor3000\relax}%
\providecommand \BibitemShut  [1]{\csname bibitem#1\endcsname}%
\let\auto@bib@innerbib\@empty
\bibitem [{\citenamefont {Bliokh}\ \emph {et~al.}(2015)\citenamefont {Bliokh},
  \citenamefont {Rodr\'{i}guez-Fortu\~{n}o}, \citenamefont {Nori},\ and\
  \citenamefont {Zayats}}]{bliokh2015spin}%
  \BibitemOpen
  \bibfield  {author} {\bibinfo {author} {\bibfnamefont {K.~Y.}\ \bibnamefont
  {Bliokh}}, \bibinfo {author} {\bibfnamefont {F.~J.}\ \bibnamefont
  {Rodr\'{i}guez-Fortu\~{n}o}}, \bibinfo {author} {\bibfnamefont
  {F.}~\bibnamefont {Nori}}, \ and\ \bibinfo {author} {\bibfnamefont {A.~V.}\
  \bibnamefont {Zayats}},\ }\bibfield  {title} {\enquote {\bibinfo {title}
  {Spin–orbit interactions of light},}\ }\href {\doibase
  10.1038/nphoton.2015.201} {\bibfield  {journal} {\bibinfo  {journal} {Nat.
  Photonics}\ }\textbf {\bibinfo {volume} {9}},\ \bibinfo {pages} {796}
  (\bibinfo {year} {2015})}\BibitemShut {NoStop}%
\bibitem [{\citenamefont {Xie}\ \emph {et~al.}(2014)\citenamefont {Xie},
  \citenamefont {Chen}, \citenamefont {Yang},\ and\ \citenamefont
  {Zhou}}]{xie2014harnessing}%
  \BibitemOpen
  \bibfield  {author} {\bibinfo {author} {\bibfnamefont {X.~S.}\ \bibnamefont
  {Xie}}, \bibinfo {author} {\bibfnamefont {Y.~Z.}\ \bibnamefont {Chen}},
  \bibinfo {author} {\bibfnamefont {K.}~\bibnamefont {Yang}}, \ and\ \bibinfo
  {author} {\bibfnamefont {J.~Y.}\ \bibnamefont {Zhou}},\ }\bibfield  {title}
  {\enquote {\bibinfo {title} {Harnessing the point-spread function for
  high-resolution far-field optical microscopy},}\ }\href {\doibase
  10.1103/PhysRevLett.113.263901} {\bibfield  {journal} {\bibinfo  {journal}
  {Phys. Rev. Lett.}\ }\textbf {\bibinfo {volume} {113}},\ \bibinfo {pages}
  {263901} (\bibinfo {year} {2014})}\BibitemShut {NoStop}%
\bibitem [{\citenamefont {Rubinsztein-Dunlop}\ \emph
  {et~al.}(2016)\citenamefont {Rubinsztein-Dunlop}, \citenamefont {Forbes},
  \citenamefont {Berry}, \citenamefont {Dennis}, \citenamefont {Andrews},
  \citenamefont {Mansuripur}, \citenamefont {Denz}, \citenamefont {Alpmann},
  \citenamefont {Banzer}, \citenamefont {Bauer} \emph
  {et~al.}}]{Rubinsztein2017Roadmap}%
  \BibitemOpen
  \bibfield  {author} {\bibinfo {author} {\bibfnamefont {H.}~\bibnamefont
  {Rubinsztein-Dunlop}}, \bibinfo {author} {\bibfnamefont {A.}~\bibnamefont
  {Forbes}}, \bibinfo {author} {\bibfnamefont {M.~V.}\ \bibnamefont {Berry}},
  \bibinfo {author} {\bibfnamefont {M.~R.}\ \bibnamefont {Dennis}}, \bibinfo
  {author} {\bibfnamefont {D.~L.}\ \bibnamefont {Andrews}}, \bibinfo {author}
  {\bibfnamefont {M.}~\bibnamefont {Mansuripur}}, \bibinfo {author}
  {\bibfnamefont {C.}~\bibnamefont {Denz}}, \bibinfo {author} {\bibfnamefont
  {C.}~\bibnamefont {Alpmann}}, \bibinfo {author} {\bibfnamefont
  {P.}~\bibnamefont {Banzer}}, \bibinfo {author} {\bibfnamefont
  {T.}~\bibnamefont {Bauer}},  \emph {et~al.},\ }\bibfield  {title} {\enquote
  {\bibinfo {title} {Roadmap on structured light},}\ }\href {\doibase
  10.1088/2040-8978/19/1/013001} {\bibfield  {journal} {\bibinfo  {journal} {J.
  Opt.}\ }\textbf {\bibinfo {volume} {19}},\ \bibinfo {pages} {013001}
  (\bibinfo {year} {2016})}\BibitemShut {NoStop}%
\bibitem [{\citenamefont {Forbes}\ \emph {et~al.}(2021)\citenamefont {Forbes},
  \citenamefont {De~Oliveira},\ and\ \citenamefont
  {Dennis}}]{forbes2021structured}%
  \BibitemOpen
  \bibfield  {author} {\bibinfo {author} {\bibfnamefont {A.}~\bibnamefont
  {Forbes}}, \bibinfo {author} {\bibfnamefont {M.}~\bibnamefont {De~Oliveira}},
  \ and\ \bibinfo {author} {\bibfnamefont {M.~R.}\ \bibnamefont {Dennis}},\
  }\bibfield  {title} {\enquote {\bibinfo {title} {Structured light},}\ }\href
  {\doibase 10.1038/s41566-021-00780-4} {\bibfield  {journal} {\bibinfo
  {journal} {Nat. Photonics}\ }\textbf {\bibinfo {volume} {15}},\ \bibinfo
  {pages} {253} (\bibinfo {year} {2021})}\BibitemShut {NoStop}%
\bibitem [{\citenamefont {Marco}\ \emph {et~al.}(2025)\citenamefont {Marco},
  \citenamefont {Mihail}, \citenamefont {John}, \citenamefont {Alexey},
  \citenamefont {C\'{e}dric}, \citenamefont {Jorge}, \citenamefont {Dustin},\
  and\ \citenamefont {Victor}}]{Marco2025Trends}%
  \BibitemOpen
  \bibfield  {author} {\bibinfo {author} {\bibfnamefont {P.}~\bibnamefont
  {Marco}}, \bibinfo {author} {\bibfnamefont {O.~C.}\ \bibnamefont {Mihail}},
  \bibinfo {author} {\bibfnamefont {P.~P.}\ \bibnamefont {John}}, \bibinfo
  {author} {\bibfnamefont {A.}~\bibnamefont {Alexey}}, \bibinfo {author}
  {\bibfnamefont {T.}~\bibnamefont {C\'{e}dric}}, \bibinfo {author}
  {\bibfnamefont {V.}~\bibnamefont {Jorge}}, \bibinfo {author} {\bibfnamefont
  {H.~F.}\ \bibnamefont {Dustin}}, \ and\ \bibinfo {author} {\bibfnamefont
  {M.}~\bibnamefont {Victor}},\ }\bibfield  {title} {\enquote {\bibinfo {title}
  {Trends in relativistic laser--matter interaction: the promises of structured
  light},}\ }\href {\doibase 10.1364/OPTICA.558754} {\bibfield  {journal}
  {\bibinfo  {journal} {Optica}\ }\textbf {\bibinfo {volume} {12}},\ \bibinfo
  {pages} {732} (\bibinfo {year} {2025})}\BibitemShut {NoStop}%
\bibitem [{\citenamefont {Yang}\ \emph {et~al.}(2021)\citenamefont {Yang},
  \citenamefont {Ren}, \citenamefont {Chen}, \citenamefont {Arita},\ and\
  \citenamefont {Rosales-Guzm{\'a}n}}]{Yang2021Optical}%
  \BibitemOpen
  \bibfield  {author} {\bibinfo {author} {\bibfnamefont {Y.~J.}\ \bibnamefont
  {Yang}}, \bibinfo {author} {\bibfnamefont {Y.~X.}\ \bibnamefont {Ren}},
  \bibinfo {author} {\bibfnamefont {M.~Z.}\ \bibnamefont {Chen}}, \bibinfo
  {author} {\bibfnamefont {Y.}~\bibnamefont {Arita}}, \ and\ \bibinfo {author}
  {\bibfnamefont {C.}~\bibnamefont {Rosales-Guzm{\'a}n}},\ }\bibfield  {title}
  {\enquote {\bibinfo {title} {{Optical trapping with structured light: a
  review}},}\ }\href {\doibase 10.1117/1.AP.3.3.034001} {\bibfield  {journal}
  {\bibinfo  {journal} {Adv. Photonics}\ }\textbf {\bibinfo {volume} {3}},\
  \bibinfo {pages} {034001} (\bibinfo {year} {2021})}\BibitemShut {NoStop}%
\bibitem [{\citenamefont {Skelton}\ \emph {et~al.}(2013)\citenamefont
  {Skelton}, \citenamefont {Sergides}, \citenamefont {Saija}, \citenamefont
  {Iat\`{i}}, \citenamefont {Marag\'{o}},\ and\ \citenamefont
  {Jones}}]{Skelton2013Trapping}%
  \BibitemOpen
  \bibfield  {author} {\bibinfo {author} {\bibfnamefont {S.~E.}\ \bibnamefont
  {Skelton}}, \bibinfo {author} {\bibfnamefont {M.}~\bibnamefont {Sergides}},
  \bibinfo {author} {\bibfnamefont {R.}~\bibnamefont {Saija}}, \bibinfo
  {author} {\bibfnamefont {M.~A.}\ \bibnamefont {Iat\`{i}}}, \bibinfo {author}
  {\bibfnamefont {O.~M.}\ \bibnamefont {Marag\'{o}}}, \ and\ \bibinfo {author}
  {\bibfnamefont {P.~H.}\ \bibnamefont {Jones}},\ }\bibfield  {title} {\enquote
  {\bibinfo {title} {Trapping volume control in optical tweezers using
  cylindrical vector beams},}\ }\href {\doibase 10.1364/OL.38.000028}
  {\bibfield  {journal} {\bibinfo  {journal} {Opt. Lett.}\ }\textbf {\bibinfo
  {volume} {38}},\ \bibinfo {pages} {28} (\bibinfo {year} {2013})}\BibitemShut
  {NoStop}%
\bibitem [{\citenamefont {Huang}\ \emph {et~al.}(2012)\citenamefont {Huang},
  \citenamefont {Guo}, \citenamefont {Li}, \citenamefont {Ling}, \citenamefont
  {Feng},\ and\ \citenamefont {Li}}]{Huang2012Optical}%
  \BibitemOpen
  \bibfield  {author} {\bibinfo {author} {\bibfnamefont {L.}~\bibnamefont
  {Huang}}, \bibinfo {author} {\bibfnamefont {H.~L.}\ \bibnamefont {Guo}},
  \bibinfo {author} {\bibfnamefont {J.~F.}\ \bibnamefont {Li}}, \bibinfo
  {author} {\bibfnamefont {L.}~\bibnamefont {Ling}}, \bibinfo {author}
  {\bibfnamefont {B.~H.}\ \bibnamefont {Feng}}, \ and\ \bibinfo {author}
  {\bibfnamefont {Z.~Y.}\ \bibnamefont {Li}},\ }\bibfield  {title} {\enquote
  {\bibinfo {title} {Optical trapping of gold nanoparticles by cylindrical
  vector beam},}\ }\href {\doibase 10.1364/OL.37.001694} {\bibfield  {journal}
  {\bibinfo  {journal} {Opt. Lett.}\ }\textbf {\bibinfo {volume} {37}},\
  \bibinfo {pages} {1694} (\bibinfo {year} {2012})}\BibitemShut {NoStop}%
\bibitem [{\citenamefont {Donato}\ \emph {et~al.}(2012)\citenamefont {Donato},
  \citenamefont {Vasi}, \citenamefont {Sayed}, \citenamefont {Jones},
  \citenamefont {Bonaccorso}, \citenamefont {Ferrari}, \citenamefont
  {Gucciardi},\ and\ \citenamefont {Marag\`{o}}}]{Donato2012Optical}%
  \BibitemOpen
  \bibfield  {author} {\bibinfo {author} {\bibfnamefont {M.~G.}\ \bibnamefont
  {Donato}}, \bibinfo {author} {\bibfnamefont {S.}~\bibnamefont {Vasi}},
  \bibinfo {author} {\bibfnamefont {R.}~\bibnamefont {Sayed}}, \bibinfo
  {author} {\bibfnamefont {P.~H.}\ \bibnamefont {Jones}}, \bibinfo {author}
  {\bibfnamefont {F.}~\bibnamefont {Bonaccorso}}, \bibinfo {author}
  {\bibfnamefont {A.~C.}\ \bibnamefont {Ferrari}}, \bibinfo {author}
  {\bibfnamefont {P.~G.}\ \bibnamefont {Gucciardi}}, \ and\ \bibinfo {author}
  {\bibfnamefont {O.~M.}\ \bibnamefont {Marag\`{o}}},\ }\bibfield  {title}
  {\enquote {\bibinfo {title} {Optical trapping of nanotubes with cylindrical
  vector beams},}\ }\href {\doibase 10.1364/OL.37.003381} {\bibfield  {journal}
  {\bibinfo  {journal} {Opt. Lett.}\ }\textbf {\bibinfo {volume} {37}},\
  \bibinfo {pages} {3381} (\bibinfo {year} {2012})}\BibitemShut {NoStop}%
\bibitem [{\citenamefont {Moradi}\ \emph {et~al.}(2019)\citenamefont {Moradi},
  \citenamefont {Shahabadi}, \citenamefont {Madadi}, \citenamefont {Karimi},\
  and\ \citenamefont {Hajizadeh}}]{Moradi2019Efficient}%
  \BibitemOpen
  \bibfield  {author} {\bibinfo {author} {\bibfnamefont {H.}~\bibnamefont
  {Moradi}}, \bibinfo {author} {\bibfnamefont {V.}~\bibnamefont {Shahabadi}},
  \bibinfo {author} {\bibfnamefont {E.}~\bibnamefont {Madadi}}, \bibinfo
  {author} {\bibfnamefont {E.}~\bibnamefont {Karimi}}, \ and\ \bibinfo {author}
  {\bibfnamefont {F.}~\bibnamefont {Hajizadeh}},\ }\bibfield  {title} {\enquote
  {\bibinfo {title} {Efficient optical trapping with cylindrical vector
  beams},}\ }\href {\doibase 10.1364/OE.27.007266} {\bibfield  {journal}
  {\bibinfo  {journal} {Opt. Express}\ }\textbf {\bibinfo {volume} {27}},\
  \bibinfo {pages} {7266} (\bibinfo {year} {2019})}\BibitemShut {NoStop}%
\bibitem [{\citenamefont {Liu}\ \emph {et~al.}(2022)\citenamefont {Liu},
  \citenamefont {Lei}, \citenamefont {Yu}, \citenamefont {Fang}, \citenamefont
  {Ma}, \citenamefont {Liu}, \citenamefont {Zheng},\ and\ \citenamefont
  {Gao}}]{Liu2022Super}%
  \BibitemOpen
  \bibfield  {author} {\bibinfo {author} {\bibfnamefont {M.}~\bibnamefont
  {Liu}}, \bibinfo {author} {\bibfnamefont {Y.~Z.}\ \bibnamefont {Lei}},
  \bibinfo {author} {\bibfnamefont {L.}~\bibnamefont {Yu}}, \bibinfo {author}
  {\bibfnamefont {X.}~\bibnamefont {Fang}}, \bibinfo {author} {\bibfnamefont
  {Y.}~\bibnamefont {Ma}}, \bibinfo {author} {\bibfnamefont {L.~X.}\
  \bibnamefont {Liu}}, \bibinfo {author} {\bibfnamefont {J.~J.}\ \bibnamefont
  {Zheng}}, \ and\ \bibinfo {author} {\bibfnamefont {P.}~\bibnamefont {Gao}},\
  }\bibfield  {title} {\enquote {\bibinfo {title} {Super-resolution optical
  microscopy using cylindrical vector beams},}\ }\href {\doibase
  doi:10.1515/nanoph-2022-0241} {\bibfield  {journal} {\bibinfo  {journal}
  {Nanophotonics}\ }\textbf {\bibinfo {volume} {11}},\ \bibinfo {pages} {3395}
  (\bibinfo {year} {2022})}\BibitemShut {NoStop}%
\bibitem [{\citenamefont {Iliadis}\ and\ \citenamefont
  {Friman-Gayer}(2021)}]{Iliadis2021Linear}%
  \BibitemOpen
  \bibfield  {author} {\bibinfo {author} {\bibfnamefont {C.}~\bibnamefont
  {Iliadis}}\ and\ \bibinfo {author} {\bibfnamefont {U.}~\bibnamefont
  {Friman-Gayer}},\ }\bibfield  {title} {\enquote {\bibinfo {title} {Linear
  polarization–direction correlations in $\gamma$-ray scattering
  experiments.}}\ }\href {\doibase 10.1140/epja/s10050-021-00472-1} {\bibfield
  {journal} {\bibinfo  {journal} {Eur. Phys. J. A}\ }\textbf {\bibinfo {volume}
  {57}},\ \bibinfo {pages} {190} (\bibinfo {year} {2021})}\BibitemShut
  {NoStop}%
\bibitem [{\citenamefont {Zilges}\ \emph {et~al.}(2022)\citenamefont {Zilges},
  \citenamefont {Balabanski}, \citenamefont {Isaak},\ and\ \citenamefont
  {Pietralla}}]{Zilges22PPNP}%
  \BibitemOpen
  \bibfield  {author} {\bibinfo {author} {\bibfnamefont {A.}~\bibnamefont
  {Zilges}}, \bibinfo {author} {\bibfnamefont {D.~L.}\ \bibnamefont
  {Balabanski}}, \bibinfo {author} {\bibfnamefont {J.}~\bibnamefont {Isaak}}, \
  and\ \bibinfo {author} {\bibfnamefont {N.}~\bibnamefont {Pietralla}},\
  }\bibfield  {title} {\enquote {\bibinfo {title} {Photonuclear reactions-from
  basic research to applications},}\ }\href {\doibase
  10.1016/j.ppnp.2021.103903} {\bibfield  {journal} {\bibinfo  {journal} {Prog.
  Part. Nucl. Phys.}\ }\textbf {\bibinfo {volume} {122}},\ \bibinfo {pages}
  {103903} (\bibinfo {year} {2022})}\BibitemShut {NoStop}%
\bibitem [{\citenamefont {Thiel}\ \emph {et~al.}(2012)\citenamefont {Thiel},
  \citenamefont {Anisovich}, \citenamefont {Bayadilov}, \citenamefont {Bantes},
  \citenamefont {Beck}, \citenamefont {Beloglazov}, \citenamefont {Bichow},
  \citenamefont {B\"ose}, \citenamefont {Brinkmann}, \citenamefont {Challand}
  \emph {et~al.}}]{Thiel2012Well}%
  \BibitemOpen
  \bibfield  {author} {\bibinfo {author} {\bibfnamefont {A.}~\bibnamefont
  {Thiel}}, \bibinfo {author} {\bibfnamefont {A.~V.}\ \bibnamefont
  {Anisovich}}, \bibinfo {author} {\bibfnamefont {D.}~\bibnamefont
  {Bayadilov}}, \bibinfo {author} {\bibfnamefont {B.}~\bibnamefont {Bantes}},
  \bibinfo {author} {\bibfnamefont {R.}~\bibnamefont {Beck}}, \bibinfo {author}
  {\bibfnamefont {Yu.}\ \bibnamefont {Beloglazov}}, \bibinfo {author}
  {\bibfnamefont {M.}~\bibnamefont {Bichow}}, \bibinfo {author} {\bibfnamefont
  {S.}~\bibnamefont {B\"ose}}, \bibinfo {author} {\bibfnamefont {K.-Th.}\
  \bibnamefont {Brinkmann}}, \bibinfo {author} {\bibfnamefont {Th.}\
  \bibnamefont {Challand}},  \emph {et~al.} (\bibinfo {collaboration}
  {CBELSA/TAPS Collaboration}),\ }\bibfield  {title} {\enquote {\bibinfo
  {title} {Well-established nucleon resonances revisited by double-polarization
  measurements},}\ }\href {\doibase 10.1103/PhysRevLett.109.102001} {\bibfield
  {journal} {\bibinfo  {journal} {Phys. Rev. Lett.}\ }\textbf {\bibinfo
  {volume} {109}},\ \bibinfo {pages} {102001} (\bibinfo {year}
  {2012})}\BibitemShut {NoStop}%
\bibitem [{\citenamefont {Gottschall}\ \emph {et~al.}(2014)\citenamefont
  {Gottschall}, \citenamefont {Anisovich}, \citenamefont {Bantes},
  \citenamefont {Bayadilov}, \citenamefont {Beck}, \citenamefont {Bichow},
  \citenamefont {B\"ose}, \citenamefont {Brinkmann}, \citenamefont {Challand},
  \citenamefont {Crede} \emph {et~al.}}]{Gottschall2014First}%
  \BibitemOpen
  \bibfield  {author} {\bibinfo {author} {\bibfnamefont {M.}~\bibnamefont
  {Gottschall}}, \bibinfo {author} {\bibfnamefont {A.~V.}\ \bibnamefont
  {Anisovich}}, \bibinfo {author} {\bibfnamefont {B.}~\bibnamefont {Bantes}},
  \bibinfo {author} {\bibfnamefont {D.}~\bibnamefont {Bayadilov}}, \bibinfo
  {author} {\bibfnamefont {R.}~\bibnamefont {Beck}}, \bibinfo {author}
  {\bibfnamefont {M.}~\bibnamefont {Bichow}}, \bibinfo {author} {\bibfnamefont
  {S.}~\bibnamefont {B\"ose}}, \bibinfo {author} {\bibfnamefont {K.-Th.}\
  \bibnamefont {Brinkmann}}, \bibinfo {author} {\bibfnamefont {Th.}\
  \bibnamefont {Challand}}, \bibinfo {author} {\bibfnamefont {V.}~\bibnamefont
  {Crede}},  \emph {et~al.} (\bibinfo {collaboration} {CBELSA/TAPS
  Collaboration}),\ }\bibfield  {title} {\enquote {\bibinfo {title} {First
  measurement of the helicity asymmetry for
  $\ensuremath{\gamma}p\ensuremath{\rightarrow}p{\ensuremath{\pi}}^{0}$ in the
  resonance region},}\ }\href {\doibase 10.1103/PhysRevLett.112.012003}
  {\bibfield  {journal} {\bibinfo  {journal} {Phys. Rev. Lett.}\ }\textbf
  {\bibinfo {volume} {112}},\ \bibinfo {pages} {012003} (\bibinfo {year}
  {2014})}\BibitemShut {NoStop}%
\bibitem [{\citenamefont {Prokhorov}\ \emph {et~al.}(2024)\citenamefont
  {Prokhorov}, \citenamefont {Yang}, \citenamefont {Ferrazzoli}, \citenamefont
  {Vink}, \citenamefont {Slane}, \citenamefont {Costa}, \citenamefont
  {Silvestri}, \citenamefont {Zhou}, \citenamefont {Bucciantini}, \citenamefont
  {Di~Marco} \emph {et~al.}}]{Prokhorov2024Evidence}%
  \BibitemOpen
  \bibfield  {author} {\bibinfo {author} {\bibfnamefont {D.~A.}\ \bibnamefont
  {Prokhorov}}, \bibinfo {author} {\bibfnamefont {Y.~J.}\ \bibnamefont {Yang}},
  \bibinfo {author} {\bibfnamefont {R.}~\bibnamefont {Ferrazzoli}}, \bibinfo
  {author} {\bibfnamefont {J.}~\bibnamefont {Vink}}, \bibinfo {author}
  {\bibfnamefont {P.}~\bibnamefont {Slane}}, \bibinfo {author} {\bibfnamefont
  {E.}~\bibnamefont {Costa}}, \bibinfo {author} {\bibfnamefont
  {S.}~\bibnamefont {Silvestri}}, \bibinfo {author} {\bibfnamefont
  {P.}~\bibnamefont {Zhou}}, \bibinfo {author} {\bibfnamefont {N.}~\bibnamefont
  {Bucciantini}}, \bibinfo {author} {\bibfnamefont {A.}~\bibnamefont
  {Di~Marco}},  \emph {et~al.},\ }\bibfield  {title} {\enquote {\bibinfo
  {title} {{Evidence for a shock-compressed magnetic field in the northwestern
  rim of Vela Jr. from X-ray polarimetry}},}\ }\href {\doibase
  10.1051/0004-6361/202452062} {\bibfield  {journal} {\bibinfo  {journal}
  {Astron. Astrophys.}\ }\textbf {\bibinfo {volume} {692}},\ \bibinfo {pages}
  {A59} (\bibinfo {year} {2024})}\BibitemShut {NoStop}%
\bibitem [{\citenamefont {Zhan}(2009)}]{zhan2009cylindrical}%
  \BibitemOpen
  \bibfield  {author} {\bibinfo {author} {\bibfnamefont {Q.~W.}\ \bibnamefont
  {Zhan}},\ }\bibfield  {title} {\enquote {\bibinfo {title} {Cylindrical vector
  beams: {From} mathematical concepts to applications},}\ }\href {\doibase
  10.1364/AOP.1.000001} {\bibfield  {journal} {\bibinfo  {journal} {Adv. Opt.
  Photonics}\ }\textbf {\bibinfo {volume} {1}},\ \bibinfo {pages} {1} (\bibinfo
  {year} {2009})}\BibitemShut {NoStop}%
\bibitem [{\citenamefont {Chen}\ \emph {et~al.}(2018)\citenamefont {Chen},
  \citenamefont {Wan},\ and\ \citenamefont {Zhan}}]{chen2018vectorial}%
  \BibitemOpen
  \bibfield  {author} {\bibinfo {author} {\bibfnamefont {J.}~\bibnamefont
  {Chen}}, \bibinfo {author} {\bibfnamefont {C.~H.}\ \bibnamefont {Wan}}, \
  and\ \bibinfo {author} {\bibfnamefont {Q.~W.}\ \bibnamefont {Zhan}},\
  }\bibfield  {title} {\enquote {\bibinfo {title} {Vectorial optical fields:
  {Recent} advances and future prospects},}\ }\href {\doibase
  https://doi.org/10.1016/j.scib.2017.12.014} {\bibfield  {journal} {\bibinfo
  {journal} {Sci. Bull.}\ }\textbf {\bibinfo {volume} {63}},\ \bibinfo {pages}
  {54} (\bibinfo {year} {2018})}\BibitemShut {NoStop}%
\bibitem [{\citenamefont {Wang}\ and\ \citenamefont
  {Liang}(2021)}]{wang2021generation}%
  \BibitemOpen
  \bibfield  {author} {\bibinfo {author} {\bibfnamefont {J.}~\bibnamefont
  {Wang}}\ and\ \bibinfo {author} {\bibfnamefont {Y.}~\bibnamefont {Liang}},\
  }\bibfield  {title} {\enquote {\bibinfo {title} {Generation and detection of
  structured light: {A} review},}\ }\href {\doibase 10.3389/fphy.2021.688284}
  {\bibfield  {journal} {\bibinfo  {journal} {Front. Phys.}\ }\textbf {\bibinfo
  {volume} {9}},\ \bibinfo {pages} {688284} (\bibinfo {year}
  {2021})}\BibitemShut {NoStop}%
\bibitem [{\citenamefont {Yonezawa}\ \emph {et~al.}(2006)\citenamefont
  {Yonezawa}, \citenamefont {Kozawa},\ and\ \citenamefont
  {Sato}}]{yonezawa2006generation}%
  \BibitemOpen
  \bibfield  {author} {\bibinfo {author} {\bibfnamefont {K.}~\bibnamefont
  {Yonezawa}}, \bibinfo {author} {\bibfnamefont {Y.}~\bibnamefont {Kozawa}}, \
  and\ \bibinfo {author} {\bibfnamefont {S.}~\bibnamefont {Sato}},\ }\bibfield
  {title} {\enquote {\bibinfo {title} {Generation of a radially polarized laser
  beam by use of the birefringence of a c-cut {Nd:YVO} 4 crystal},}\ }\href
  {\doibase 10.1364/OL.31.002151} {\bibfield  {journal} {\bibinfo  {journal}
  {Opt. Lett.}\ }\textbf {\bibinfo {volume} {31}},\ \bibinfo {pages} {2151}
  (\bibinfo {year} {2006})}\BibitemShut {NoStop}%
\bibitem [{\citenamefont {Machavariani}\ \emph {et~al.}(2007)\citenamefont
  {Machavariani}, \citenamefont {Lumer}, \citenamefont {Moshe}, \citenamefont
  {Meir}, \citenamefont {Jackel},\ and\ \citenamefont
  {Davidson}}]{machavariani2007birefringence}%
  \BibitemOpen
  \bibfield  {author} {\bibinfo {author} {\bibfnamefont {G.}~\bibnamefont
  {Machavariani}}, \bibinfo {author} {\bibfnamefont {Y.}~\bibnamefont {Lumer}},
  \bibinfo {author} {\bibfnamefont {I.}~\bibnamefont {Moshe}}, \bibinfo
  {author} {\bibfnamefont {A.}~\bibnamefont {Meir}}, \bibinfo {author}
  {\bibfnamefont {S.}~\bibnamefont {Jackel}}, \ and\ \bibinfo {author}
  {\bibfnamefont {N.}~\bibnamefont {Davidson}},\ }\bibfield  {title} {\enquote
  {\bibinfo {title} {Birefringence-induced bifocusing for selection of radially
  or azimuthally polarized laser modes},}\ }\href {\doibase
  10.1364/AO.46.003304} {\bibfield  {journal} {\bibinfo  {journal} {Appl.
  Opt.}\ }\textbf {\bibinfo {volume} {46}},\ \bibinfo {pages} {3304} (\bibinfo
  {year} {2007})}\BibitemShut {NoStop}%
\bibitem [{\citenamefont {Bisson}\ \emph {et~al.}(2006)\citenamefont {Bisson},
  \citenamefont {Li}, \citenamefont {Ueda},\ and\ \citenamefont
  {Senatsky}}]{bisson2006radially}%
  \BibitemOpen
  \bibfield  {author} {\bibinfo {author} {\bibfnamefont {J.~F.}\ \bibnamefont
  {Bisson}}, \bibinfo {author} {\bibfnamefont {J.~L.}\ \bibnamefont {Li}},
  \bibinfo {author} {\bibfnamefont {K.}~\bibnamefont {Ueda}}, \ and\ \bibinfo
  {author} {\bibfnamefont {Y.}~\bibnamefont {Senatsky}},\ }\bibfield  {title}
  {\enquote {\bibinfo {title} {Radially polarized ring and arc beams of a
  neodymium laser with an intra-cavity axicon},}\ }\href {\doibase
  10.1364/OE.14.003304} {\bibfield  {journal} {\bibinfo  {journal} {Opt.
  Express}\ }\textbf {\bibinfo {volume} {14}},\ \bibinfo {pages} {3304}
  (\bibinfo {year} {2006})}\BibitemShut {NoStop}%
\bibitem [{\citenamefont {S{\'a}nchez-L{\'o}pez}\ \emph
  {et~al.}(2015)\citenamefont {S{\'a}nchez-L{\'o}pez}, \citenamefont {Davis},
  \citenamefont {Hashimoto}, \citenamefont {Moreno}, \citenamefont {Hurtado},
  \citenamefont {Badham}, \citenamefont {Tanabe},\ and\ \citenamefont
  {Delaney}}]{sanchez2015performance}%
  \BibitemOpen
  \bibfield  {author} {\bibinfo {author} {\bibfnamefont {M.~M.}\ \bibnamefont
  {S{\'a}nchez-L{\'o}pez}}, \bibinfo {author} {\bibfnamefont {J.~A.}\
  \bibnamefont {Davis}}, \bibinfo {author} {\bibfnamefont {N.}~\bibnamefont
  {Hashimoto}}, \bibinfo {author} {\bibfnamefont {I.}~\bibnamefont {Moreno}},
  \bibinfo {author} {\bibfnamefont {E.}~\bibnamefont {Hurtado}}, \bibinfo
  {author} {\bibfnamefont {K.}~\bibnamefont {Badham}}, \bibinfo {author}
  {\bibfnamefont {A.}~\bibnamefont {Tanabe}}, \ and\ \bibinfo {author}
  {\bibfnamefont {S.~W.}\ \bibnamefont {Delaney}},\ }\bibfield  {title}
  {\enquote {\bibinfo {title} {Performance of a q-plate tunable retarder in
  reflection for the switchable generation of both first-and second-order
  vector beams},}\ }\href {\doibase 10.1364/OL.41.000013} {\bibfield  {journal}
  {\bibinfo  {journal} {Opt. Lett.}\ }\textbf {\bibinfo {volume} {41}},\
  \bibinfo {pages} {13} (\bibinfo {year} {2015})}\BibitemShut {NoStop}%
\bibitem [{\citenamefont {Naidoo}\ \emph {et~al.}(2016)\citenamefont {Naidoo},
  \citenamefont {Roux}, \citenamefont {Dudley}, \citenamefont {Litvin},
  \citenamefont {Piccirillo}, \citenamefont {Marrucci},\ and\ \citenamefont
  {Forbes}}]{naidoo2016controlled}%
  \BibitemOpen
  \bibfield  {author} {\bibinfo {author} {\bibfnamefont {D.}~\bibnamefont
  {Naidoo}}, \bibinfo {author} {\bibfnamefont {F.~S.}\ \bibnamefont {Roux}},
  \bibinfo {author} {\bibfnamefont {A.}~\bibnamefont {Dudley}}, \bibinfo
  {author} {\bibfnamefont {I.}~\bibnamefont {Litvin}}, \bibinfo {author}
  {\bibfnamefont {B.}~\bibnamefont {Piccirillo}}, \bibinfo {author}
  {\bibfnamefont {L.}~\bibnamefont {Marrucci}}, \ and\ \bibinfo {author}
  {\bibfnamefont {A.}~\bibnamefont {Forbes}},\ }\bibfield  {title} {\enquote
  {\bibinfo {title} {Controlled generation of higher-order {Poincar{\'e}}
  sphere beams from a laser},}\ }\href {\doibase 10.1038/nphoton.2016.37}
  {\bibfield  {journal} {\bibinfo  {journal} {Nat. Photonics}\ }\textbf
  {\bibinfo {volume} {10}},\ \bibinfo {pages} {327} (\bibinfo {year}
  {2016})}\BibitemShut {NoStop}%
\bibitem [{\citenamefont {Sun}\ \emph {et~al.}(2012)\citenamefont {Sun},
  \citenamefont {Wang}, \citenamefont {Xu}, \citenamefont {Gu}, \citenamefont
  {Lin}, \citenamefont {Ming},\ and\ \citenamefont {Zhan}}]{sun2012low}%
  \BibitemOpen
  \bibfield  {author} {\bibinfo {author} {\bibfnamefont {B.}~\bibnamefont
  {Sun}}, \bibinfo {author} {\bibfnamefont {A.}~\bibnamefont {Wang}}, \bibinfo
  {author} {\bibfnamefont {L.~X.}\ \bibnamefont {Xu}}, \bibinfo {author}
  {\bibfnamefont {C.}~\bibnamefont {Gu}}, \bibinfo {author} {\bibfnamefont
  {Z.~X.}\ \bibnamefont {Lin}}, \bibinfo {author} {\bibfnamefont
  {H.}~\bibnamefont {Ming}}, \ and\ \bibinfo {author} {\bibfnamefont {Q.~W.}\
  \bibnamefont {Zhan}},\ }\bibfield  {title} {\enquote {\bibinfo {title}
  {{Low-threshold single-wavelength all-fiber laser generating cylindrical
  vector beams using a few-mode fiber Bragg grating}},}\ }\href {\doibase
  10.1364/OL.37.000464} {\bibfield  {journal} {\bibinfo  {journal} {Opt.
  Lett.}\ }\textbf {\bibinfo {volume} {37}},\ \bibinfo {pages} {464} (\bibinfo
  {year} {2012})}\BibitemShut {NoStop}%
\bibitem [{\citenamefont {Lin}\ \emph {et~al.}(2015)\citenamefont {Lin},
  \citenamefont {Yan}, \citenamefont {Zhou}, \citenamefont {Xu}, \citenamefont
  {Gu},\ and\ \citenamefont {Zhan}}]{lin2015tungsten}%
  \BibitemOpen
  \bibfield  {author} {\bibinfo {author} {\bibfnamefont {J.}~\bibnamefont
  {Lin}}, \bibinfo {author} {\bibfnamefont {K.}~\bibnamefont {Yan}}, \bibinfo
  {author} {\bibfnamefont {Y.}~\bibnamefont {Zhou}}, \bibinfo {author}
  {\bibfnamefont {L.~X.}\ \bibnamefont {Xu}}, \bibinfo {author} {\bibfnamefont
  {C.}~\bibnamefont {Gu}}, \ and\ \bibinfo {author} {\bibfnamefont {Q.~W.}\
  \bibnamefont {Zhan}},\ }\bibfield  {title} {\enquote {\bibinfo {title}
  {Tungsten disulphide based all fiber {Q}-switching cylindrical-vector beam
  generation},}\ }\href {https://doi.org/10.1063/1.4935465} {\bibfield
  {journal} {\bibinfo  {journal} {Appl. Phys. Lett.}\ }\textbf {\bibinfo
  {volume} {107}} (\bibinfo {year} {2015})}\BibitemShut {NoStop}%
\bibitem [{\citenamefont {Mao}\ \emph {et~al.}(2021)\citenamefont {Mao},
  \citenamefont {Zheng}, \citenamefont {Zeng}, \citenamefont {Lu},
  \citenamefont {Wang}, \citenamefont {Zhang}, \citenamefont {Zhang},
  \citenamefont {Mei},\ and\ \citenamefont {Zhao}}]{mao2021generation}%
  \BibitemOpen
  \bibfield  {author} {\bibinfo {author} {\bibfnamefont {D.}~\bibnamefont
  {Mao}}, \bibinfo {author} {\bibfnamefont {Y.}~\bibnamefont {Zheng}}, \bibinfo
  {author} {\bibfnamefont {C.}~\bibnamefont {Zeng}}, \bibinfo {author}
  {\bibfnamefont {H.}~\bibnamefont {Lu}}, \bibinfo {author} {\bibfnamefont
  {C.}~\bibnamefont {Wang}}, \bibinfo {author} {\bibfnamefont {H.}~\bibnamefont
  {Zhang}}, \bibinfo {author} {\bibfnamefont {W.}~\bibnamefont {Zhang}},
  \bibinfo {author} {\bibfnamefont {T.}~\bibnamefont {Mei}}, \ and\ \bibinfo
  {author} {\bibfnamefont {J.~L.}\ \bibnamefont {Zhao}},\ }\bibfield  {title}
  {\enquote {\bibinfo {title} {Generation of polarization and phase singular
  beams in fibers and fiber lasers},}\ }\href {\doibase
  10.1117/1.AP.3.1.014002} {\bibfield  {journal} {\bibinfo  {journal} {Adv.
  Photonics}\ }\textbf {\bibinfo {volume} {3}},\ \bibinfo {pages} {014002}
  (\bibinfo {year} {2021})}\BibitemShut {NoStop}%
\bibitem [{\citenamefont {Mark}\ \emph {et~al.}(2002)\citenamefont {Mark},
  \citenamefont {Farnaz}, \citenamefont {Rimvydas},\ and\ \citenamefont
  {Tony}}]{neil2002method}%
  \BibitemOpen
  \bibfield  {author} {\bibinfo {author} {\bibfnamefont {A.~A.~N.}\
  \bibnamefont {Mark}}, \bibinfo {author} {\bibfnamefont {M.}~\bibnamefont
  {Farnaz}}, \bibinfo {author} {\bibfnamefont {J.}~\bibnamefont {Rimvydas}}, \
  and\ \bibinfo {author} {\bibfnamefont {W.}~\bibnamefont {Tony}},\ }\bibfield
  {title} {\enquote {\bibinfo {title} {Method for the generation of arbitrary
  complex vector wave fronts},}\ }\href {\doibase 10.1364/OL.27.001929}
  {\bibfield  {journal} {\bibinfo  {journal} {Opt. Lett.}\ }\textbf {\bibinfo
  {volume} {27}},\ \bibinfo {pages} {1929} (\bibinfo {year}
  {2002})}\BibitemShut {NoStop}%
\bibitem [{\citenamefont {Maurer}\ \emph {et~al.}(2007)\citenamefont {Maurer},
  \citenamefont {Jesacher}, \citenamefont {F{\"u}rhapter}, \citenamefont
  {Bernet},\ and\ \citenamefont {Ritsch-Marte}}]{maurer2007tailoring}%
  \BibitemOpen
  \bibfield  {author} {\bibinfo {author} {\bibfnamefont {C.}~\bibnamefont
  {Maurer}}, \bibinfo {author} {\bibfnamefont {A.}~\bibnamefont {Jesacher}},
  \bibinfo {author} {\bibfnamefont {S.}~\bibnamefont {F{\"u}rhapter}}, \bibinfo
  {author} {\bibfnamefont {S.}~\bibnamefont {Bernet}}, \ and\ \bibinfo {author}
  {\bibfnamefont {M.}~\bibnamefont {Ritsch-Marte}},\ }\bibfield  {title}
  {\enquote {\bibinfo {title} {Tailoring of arbitrary optical vector beams},}\
  }\href {\doibase 10.1088/1367-2630/9/3/078} {\bibfield  {journal} {\bibinfo
  {journal} {New J. Phys.}\ }\textbf {\bibinfo {volume} {9}},\ \bibinfo {pages}
  {78} (\bibinfo {year} {2007})}\BibitemShut {NoStop}%
\bibitem [{\citenamefont {Ruiz}\ \emph {et~al.}(2013)\citenamefont {Ruiz},
  \citenamefont {Pagliusi}, \citenamefont {Provenzano},\ and\ \citenamefont
  {Cipparrone}}]{ruiz2013highly}%
  \BibitemOpen
  \bibfield  {author} {\bibinfo {author} {\bibfnamefont {U.}~\bibnamefont
  {Ruiz}}, \bibinfo {author} {\bibfnamefont {P.}~\bibnamefont {Pagliusi}},
  \bibinfo {author} {\bibfnamefont {C.}~\bibnamefont {Provenzano}}, \ and\
  \bibinfo {author} {\bibfnamefont {G.}~\bibnamefont {Cipparrone}},\ }\bibfield
   {title} {\enquote {\bibinfo {title} {Highly efficient generation of vector
  beams through polarization holograms},}\ }\href {\doibase 10.1063/1.4801317}
  {\bibfield  {journal} {\bibinfo  {journal} {Appl. Phys. Lett.}\ }\textbf
  {\bibinfo {volume} {102}},\ \bibinfo {pages} {161104} (\bibinfo {year}
  {2013})}\BibitemShut {NoStop}%
\bibitem [{\citenamefont {Han}\ \emph {et~al.}(2013)\citenamefont {Han},
  \citenamefont {Yang}, \citenamefont {Cheng},\ and\ \citenamefont
  {Zhan}}]{han2013vectorial}%
  \BibitemOpen
  \bibfield  {author} {\bibinfo {author} {\bibfnamefont {W.}~\bibnamefont
  {Han}}, \bibinfo {author} {\bibfnamefont {Y.~F.}\ \bibnamefont {Yang}},
  \bibinfo {author} {\bibfnamefont {W.}~\bibnamefont {Cheng}}, \ and\ \bibinfo
  {author} {\bibfnamefont {Q.~W.}\ \bibnamefont {Zhan}},\ }\bibfield  {title}
  {\enquote {\bibinfo {title} {Vectorial optical field generator for the
  creation of arbitrarily complex fields},}\ }\href {\doibase
  10.1364/OE.21.020692} {\bibfield  {journal} {\bibinfo  {journal} {Opt.
  Express}\ }\textbf {\bibinfo {volume} {21}},\ \bibinfo {pages} {20692}
  (\bibinfo {year} {2013})}\BibitemShut {NoStop}%
\bibitem [{\citenamefont {Kevin}\ \emph {et~al.}(2016)\citenamefont {Kevin},
  \citenamefont {Sergey}, \citenamefont {Miles}, \citenamefont
  {Tom\'{a}\v{s}},\ and\ \citenamefont {David}}]{mitchell2016high}%
  \BibitemOpen
  \bibfield  {author} {\bibinfo {author} {\bibfnamefont {J.~M.}\ \bibnamefont
  {Kevin}}, \bibinfo {author} {\bibfnamefont {T.}~\bibnamefont {Sergey}},
  \bibinfo {author} {\bibfnamefont {J.~P.}\ \bibnamefont {Miles}}, \bibinfo
  {author} {\bibfnamefont {\v{C}.}\ \bibnamefont {Tom\'{a}\v{s}}}, \ and\
  \bibinfo {author} {\bibfnamefont {B.~P.}\ \bibnamefont {David}},\ }\bibfield
  {title} {\enquote {\bibinfo {title} {High-speed spatial control of the
  intensity, phase and polarisation of vector beams using a digital
  micro-mirror device},}\ }\href {\doibase 10.1364/OE.24.029269} {\bibfield
  {journal} {\bibinfo  {journal} {Opt. Express}\ }\textbf {\bibinfo {volume}
  {24}},\ \bibinfo {pages} {29269} (\bibinfo {year} {2016})}\BibitemShut
  {NoStop}%
\bibitem [{\citenamefont {Hirayama}\ \emph {et~al.}(2006)\citenamefont
  {Hirayama}, \citenamefont {Kozawa}, \citenamefont {Nakamura},\ and\
  \citenamefont {Sato}}]{hirayama2006generation}%
  \BibitemOpen
  \bibfield  {author} {\bibinfo {author} {\bibfnamefont {T.}~\bibnamefont
  {Hirayama}}, \bibinfo {author} {\bibfnamefont {Y.}~\bibnamefont {Kozawa}},
  \bibinfo {author} {\bibfnamefont {T.}~\bibnamefont {Nakamura}}, \ and\
  \bibinfo {author} {\bibfnamefont {S.}~\bibnamefont {Sato}},\ }\bibfield
  {title} {\enquote {\bibinfo {title} {Generation of a cylindrically symmetric,
  polarized laser beam with narrow linewidth and fine tunability},}\ }\href
  {\doibase 10.1364/OE.14.012839} {\bibfield  {journal} {\bibinfo  {journal}
  {Opt. Express}\ }\textbf {\bibinfo {volume} {14}},\ \bibinfo {pages} {12839}
  (\bibinfo {year} {2006})}\BibitemShut {NoStop}%
\bibitem [{\citenamefont {Liu}\ \emph {et~al.}(2014)\citenamefont {Liu},
  \citenamefont {Ling}, \citenamefont {Yi}, \citenamefont {Zhou}, \citenamefont
  {Luo},\ and\ \citenamefont {Wen}}]{liu2014realization}%
  \BibitemOpen
  \bibfield  {author} {\bibinfo {author} {\bibfnamefont {Y.~C.}\ \bibnamefont
  {Liu}}, \bibinfo {author} {\bibfnamefont {X.~H.}\ \bibnamefont {Ling}},
  \bibinfo {author} {\bibfnamefont {X.~N.}\ \bibnamefont {Yi}}, \bibinfo
  {author} {\bibfnamefont {X.~X.}\ \bibnamefont {Zhou}}, \bibinfo {author}
  {\bibfnamefont {H.~L.}\ \bibnamefont {Luo}}, \ and\ \bibinfo {author}
  {\bibfnamefont {S.~C.}\ \bibnamefont {Wen}},\ }\bibfield  {title} {\enquote
  {\bibinfo {title} {Realization of polarization evolution on higher-order
  {Poincar{\'e}} sphere with metasurface},}\ }\href {\doibase
  10.1063/1.4878409} {\bibfield  {journal} {\bibinfo  {journal} {Appl. Phys.
  Lett.}\ }\textbf {\bibinfo {volume} {104}} (\bibinfo {year} {2014}),\
  10.1063/1.4878409}\BibitemShut {NoStop}%
\bibitem [{\citenamefont {Yue}\ \emph {et~al.}(2016)\citenamefont {Yue},
  \citenamefont {Wen}, \citenamefont {Xin}, \citenamefont {Gerardot},
  \citenamefont {Li},\ and\ \citenamefont {Chen}}]{yue2016vector}%
  \BibitemOpen
  \bibfield  {author} {\bibinfo {author} {\bibfnamefont {F.~Y.}\ \bibnamefont
  {Yue}}, \bibinfo {author} {\bibfnamefont {D.~D.}\ \bibnamefont {Wen}},
  \bibinfo {author} {\bibfnamefont {J.~T.}\ \bibnamefont {Xin}}, \bibinfo
  {author} {\bibfnamefont {B.~D.}\ \bibnamefont {Gerardot}}, \bibinfo {author}
  {\bibfnamefont {J.~S.}\ \bibnamefont {Li}}, \ and\ \bibinfo {author}
  {\bibfnamefont {X.~Z.}\ \bibnamefont {Chen}},\ }\bibfield  {title} {\enquote
  {\bibinfo {title} {Vector vortex beam generation with a single plasmonic
  metasurface},}\ }\href {\doibase 10.1021/acsphotonics.6b00392} {\bibfield
  {journal} {\bibinfo  {journal} {ACS Photonics}\ }\textbf {\bibinfo {volume}
  {3}},\ \bibinfo {pages} {1558} (\bibinfo {year} {2016})}\BibitemShut
  {NoStop}%
\bibitem [{\citenamefont {Zhao}\ \emph {et~al.}(2025)\citenamefont {Zhao},
  \citenamefont {Jing},\ and\ \citenamefont {Yu}}]{zhao2025research}%
  \BibitemOpen
  \bibfield  {author} {\bibinfo {author} {\bibfnamefont {F.~F.}\ \bibnamefont
  {Zhao}}, \bibinfo {author} {\bibfnamefont {X.~F.}\ \bibnamefont {Jing}}, \
  and\ \bibinfo {author} {\bibfnamefont {M.~Z.}\ \bibnamefont {Yu}},\
  }\bibfield  {title} {\enquote {\bibinfo {title} {Research progress on the
  principle and application of metalenses based on metasurfaces},}\ }\href
  {\doibase 10.1063/5.0246029} {\bibfield  {journal} {\bibinfo  {journal} {J.
  Appl. Phys.}\ }\textbf {\bibinfo {volume} {137}},\ \bibinfo {pages} {050701}
  (\bibinfo {year} {2025})}\BibitemShut {NoStop}%
\bibitem [{\citenamefont {Olsen}\ and\ \citenamefont
  {Maximon}(1959)}]{olsen1959photon}%
  \BibitemOpen
  \bibfield  {author} {\bibinfo {author} {\bibfnamefont {H.}~\bibnamefont
  {Olsen}}\ and\ \bibinfo {author} {\bibfnamefont {L.~C.}\ \bibnamefont
  {Maximon}},\ }\bibfield  {title} {\enquote {\bibinfo {title} {Photon and
  electron polarization in high-energy bremsstrahlung and pair production with
  screening},}\ }\href {\doibase 10.1103/PhysRev.114.887} {\bibfield  {journal}
  {\bibinfo  {journal} {Phys. Rev.}\ }\textbf {\bibinfo {volume} {114}},\
  \bibinfo {pages} {887} (\bibinfo {year} {1959})}\BibitemShut {NoStop}%
\bibitem [{\citenamefont {Kuraev}\ \emph {et~al.}(2010)\citenamefont {Kuraev},
  \citenamefont {Bystritskiy}, \citenamefont {Shatnev},\ and\ \citenamefont
  {Tomasi-Gustafsson}}]{Kuraev2010Bremsstrahlung}%
  \BibitemOpen
  \bibfield  {author} {\bibinfo {author} {\bibfnamefont {E.~A.}\ \bibnamefont
  {Kuraev}}, \bibinfo {author} {\bibfnamefont {Yu.~M.}\ \bibnamefont
  {Bystritskiy}}, \bibinfo {author} {\bibfnamefont {M.}~\bibnamefont
  {Shatnev}}, \ and\ \bibinfo {author} {\bibfnamefont {E.}~\bibnamefont
  {Tomasi-Gustafsson}},\ }\bibfield  {title} {\enquote {\bibinfo {title}
  {Bremsstrahlung and pair production processes at low energies:
  Multidifferential cross section and polarization phenomena},}\ }\href
  {\doibase 10.1103/PhysRevC.81.055208} {\bibfield  {journal} {\bibinfo
  {journal} {Phys. Rev. C}\ }\textbf {\bibinfo {volume} {81}},\ \bibinfo
  {pages} {055208} (\bibinfo {year} {2010})}\BibitemShut {NoStop}%
\bibitem [{\citenamefont {Howell}\ \emph {et~al.}(2021)\citenamefont {Howell},
  \citenamefont {Ahmed}, \citenamefont {Afanasev}, \citenamefont {Alesini},
  \citenamefont {Annand}, \citenamefont {Aprahamian}, \citenamefont
  {Balabanski}, \citenamefont {Benson}, \citenamefont {Bernstein},
  \citenamefont {Brune} \emph {et~al.}}]{Howell2021International}%
  \BibitemOpen
  \bibfield  {author} {\bibinfo {author} {\bibfnamefont {C.~R.}\ \bibnamefont
  {Howell}}, \bibinfo {author} {\bibfnamefont {M.~W.}\ \bibnamefont {Ahmed}},
  \bibinfo {author} {\bibfnamefont {A.}~\bibnamefont {Afanasev}}, \bibinfo
  {author} {\bibfnamefont {D.}~\bibnamefont {Alesini}}, \bibinfo {author}
  {\bibfnamefont {J.~R.~M.}\ \bibnamefont {Annand}}, \bibinfo {author}
  {\bibfnamefont {A.}~\bibnamefont {Aprahamian}}, \bibinfo {author}
  {\bibfnamefont {D.~L.}\ \bibnamefont {Balabanski}}, \bibinfo {author}
  {\bibfnamefont {S.~V.}\ \bibnamefont {Benson}}, \bibinfo {author}
  {\bibfnamefont {A.}~\bibnamefont {Bernstein}}, \bibinfo {author}
  {\bibfnamefont {C.~R.}\ \bibnamefont {Brune}},  \emph {et~al.},\ }\bibfield
  {title} {\enquote {\bibinfo {title} {International workshop on next
  generation gamma-ray source},}\ }\href {\doibase 10.1088/1361-6471/ac2827}
  {\bibfield  {journal} {\bibinfo  {journal} {J. Phys. G: Nucl. Part. Phys.}\
  }\textbf {\bibinfo {volume} {49}},\ \bibinfo {pages} {010502} (\bibinfo
  {year} {2021})}\BibitemShut {NoStop}%
\bibitem [{\citenamefont {Li}\ \emph {et~al.}(2020)\citenamefont {Li},
  \citenamefont {Shaisultanov}, \citenamefont {Chen}, \citenamefont {Wan},
  \citenamefont {Hatsagortsyan}, \citenamefont {Keitel},\ and\ \citenamefont
  {Li}}]{li2020polarized}%
  \BibitemOpen
  \bibfield  {author} {\bibinfo {author} {\bibfnamefont {Y.~F.}\ \bibnamefont
  {Li}}, \bibinfo {author} {\bibfnamefont {R.}~\bibnamefont {Shaisultanov}},
  \bibinfo {author} {\bibfnamefont {Y.~Y.}\ \bibnamefont {Chen}}, \bibinfo
  {author} {\bibfnamefont {F.}~\bibnamefont {Wan}}, \bibinfo {author}
  {\bibfnamefont {K.~Z.}\ \bibnamefont {Hatsagortsyan}}, \bibinfo {author}
  {\bibfnamefont {C.~H.}\ \bibnamefont {Keitel}}, \ and\ \bibinfo {author}
  {\bibfnamefont {J.~X.}\ \bibnamefont {Li}},\ }\bibfield  {title} {\enquote
  {\bibinfo {title} {Polarized ultrashort brilliant multi-{GeV}
  $\ensuremath{\gamma}$ rays via single-shot laser-electron interaction},}\
  }\href {\doibase 10.1103/PhysRevLett.124.014801} {\bibfield  {journal}
  {\bibinfo  {journal} {Phys. Rev. Lett.}\ }\textbf {\bibinfo {volume} {124}},\
  \bibinfo {pages} {014801} (\bibinfo {year} {2020})}\BibitemShut {NoStop}%
\bibitem [{\citenamefont {Xue}\ \emph {et~al.}(2020)\citenamefont {Xue},
  \citenamefont {Dou}, \citenamefont {Wan}, \citenamefont {Yu}, \citenamefont
  {Wang}, \citenamefont {Ren}, \citenamefont {Zhao}, \citenamefont {Zhao},
  \citenamefont {Xu},\ and\ \citenamefont {Li}}]{Xue2020Generation}%
  \BibitemOpen
  \bibfield  {author} {\bibinfo {author} {\bibfnamefont {K.}~\bibnamefont
  {Xue}}, \bibinfo {author} {\bibfnamefont {Z.~K.}\ \bibnamefont {Dou}},
  \bibinfo {author} {\bibfnamefont {F.}~\bibnamefont {Wan}}, \bibinfo {author}
  {\bibfnamefont {T.~P.}\ \bibnamefont {Yu}}, \bibinfo {author} {\bibfnamefont
  {W.~M.}\ \bibnamefont {Wang}}, \bibinfo {author} {\bibfnamefont {J.~R.}\
  \bibnamefont {Ren}}, \bibinfo {author} {\bibfnamefont {Q.}~\bibnamefont
  {Zhao}}, \bibinfo {author} {\bibfnamefont {Y.~T.}\ \bibnamefont {Zhao}},
  \bibinfo {author} {\bibfnamefont {Z.~F.}\ \bibnamefont {Xu}}, \ and\ \bibinfo
  {author} {\bibfnamefont {J.~X.}\ \bibnamefont {Li}},\ }\bibfield  {title}
  {\enquote {\bibinfo {title} {Generation of highly-polarized high-energy
  brilliant $\gamma$-rays via laser-plasma interaction},}\ }\href {\doibase
  10.1063/5.0007734} {\bibfield  {journal} {\bibinfo  {journal} {Matter Radiat.
  Extremes}\ }\textbf {\bibinfo {volume} {5}},\ \bibinfo {pages} {054402}
  (\bibinfo {year} {2020})}\BibitemShut {NoStop}%
\bibitem [{\citenamefont {Wang}\ \emph {et~al.}(2024)\citenamefont {Wang},
  \citenamefont {Ababekri}, \citenamefont {Wan}, \citenamefont {Wen},
  \citenamefont {Wei}, \citenamefont {Li}, \citenamefont {Kang}, \citenamefont
  {Zhang}, \citenamefont {Zhao}, \citenamefont {Zhou} \emph
  {et~al.}}]{wang2024manipulation}%
  \BibitemOpen
  \bibfield  {author} {\bibinfo {author} {\bibfnamefont {Y.}~\bibnamefont
  {Wang}}, \bibinfo {author} {\bibfnamefont {M.}~\bibnamefont {Ababekri}},
  \bibinfo {author} {\bibfnamefont {F.}~\bibnamefont {Wan}}, \bibinfo {author}
  {\bibfnamefont {J.~X.}\ \bibnamefont {Wen}}, \bibinfo {author} {\bibfnamefont
  {W.~Q.}\ \bibnamefont {Wei}}, \bibinfo {author} {\bibfnamefont {Z.~P.}\
  \bibnamefont {Li}}, \bibinfo {author} {\bibfnamefont {H.~T.}\ \bibnamefont
  {Kang}}, \bibinfo {author} {\bibfnamefont {B.}~\bibnamefont {Zhang}},
  \bibinfo {author} {\bibfnamefont {Y.~T.}\ \bibnamefont {Zhao}}, \bibinfo
  {author} {\bibfnamefont {W.~M.}\ \bibnamefont {Zhou}},  \emph {et~al.},\
  }\bibfield  {title} {\enquote {\bibinfo {title} {Manipulation of $\gamma$-ray
  polarization in {Compton} scattering},}\ }\href
  {https://doi.org/10.1063/5.0191466} {\bibfield  {journal} {\bibinfo
  {journal} {Phys. Plasmas}\ }\textbf {\bibinfo {volume} {31}} (\bibinfo {year}
  {2024})}\BibitemShut {NoStop}%
\bibitem [{\citenamefont {Giulietti}\ \emph {et~al.}(2008)\citenamefont
  {Giulietti}, \citenamefont {Bourgeois}, \citenamefont {Ceccotti},
  \citenamefont {Davoine}, \citenamefont {Dobosz}, \citenamefont {D'Oliveira},
  \citenamefont {Galimberti}, \citenamefont {Galy}, \citenamefont {Gamucci},
  \citenamefont {Giulietti} \emph {et~al.}}]{Giulietti2008Intense}%
  \BibitemOpen
  \bibfield  {author} {\bibinfo {author} {\bibfnamefont {A.}~\bibnamefont
  {Giulietti}}, \bibinfo {author} {\bibfnamefont {N.}~\bibnamefont
  {Bourgeois}}, \bibinfo {author} {\bibfnamefont {T.}~\bibnamefont {Ceccotti}},
  \bibinfo {author} {\bibfnamefont {X.}~\bibnamefont {Davoine}}, \bibinfo
  {author} {\bibfnamefont {S.}~\bibnamefont {Dobosz}}, \bibinfo {author}
  {\bibfnamefont {P.}~\bibnamefont {D'Oliveira}}, \bibinfo {author}
  {\bibfnamefont {M.}~\bibnamefont {Galimberti}}, \bibinfo {author}
  {\bibfnamefont {J.}~\bibnamefont {Galy}}, \bibinfo {author} {\bibfnamefont
  {A.}~\bibnamefont {Gamucci}}, \bibinfo {author} {\bibfnamefont
  {D.}~\bibnamefont {Giulietti}},  \emph {et~al.},\ }\bibfield  {title}
  {\enquote {\bibinfo {title} {Intense $\ensuremath{\gamma}$-ray source in the
  giant-dipole-resonance range driven by {10-TW} laser pulses},}\ }\href
  {\doibase 10.1103/PhysRevLett.101.105002} {\bibfield  {journal} {\bibinfo
  {journal} {Phys. Rev. Lett.}\ }\textbf {\bibinfo {volume} {101}},\ \bibinfo
  {pages} {105002} (\bibinfo {year} {2008})}\BibitemShut {NoStop}%
\bibitem [{\citenamefont {Albert}\ and\ \citenamefont
  {Thomas}(2016)}]{albert2016applications}%
  \BibitemOpen
  \bibfield  {author} {\bibinfo {author} {\bibfnamefont {F.}~\bibnamefont
  {Albert}}\ and\ \bibinfo {author} {\bibfnamefont {A.~G.~R.}\ \bibnamefont
  {Thomas}},\ }\bibfield  {title} {\enquote {\bibinfo {title} {Applications of
  laser wakefield accelerator-based light sources},}\ }\href {\doibase
  10.1088/0741-3335/58/10/103001} {\bibfield  {journal} {\bibinfo  {journal}
  {Plasma Phys. Control. Fusion}\ }\textbf {\bibinfo {volume} {58}},\ \bibinfo
  {pages} {103001} (\bibinfo {year} {2016})}\BibitemShut {NoStop}%
\bibitem [{\citenamefont {Uggerh\o{}j}(2005)}]{Uggerhoj2005The}%
  \BibitemOpen
  \bibfield  {author} {\bibinfo {author} {\bibfnamefont {U.~I.}\ \bibnamefont
  {Uggerh\o{}j}},\ }\bibfield  {title} {\enquote {\bibinfo {title} {The
  interaction of relativistic particles with strong crystalline fields},}\
  }\href {\doibase 10.1103/RevModPhys.77.1131} {\bibfield  {journal} {\bibinfo
  {journal} {Rev. Mod. Phys.}\ }\textbf {\bibinfo {volume} {77}},\ \bibinfo
  {pages} {1131} (\bibinfo {year} {2005})}\BibitemShut {NoStop}%
\bibitem [{\citenamefont {Ter-Mikaelian}(1972)}]{ter1972high}%
  \BibitemOpen
  \bibfield  {author} {\bibinfo {author} {\bibfnamefont {M.~L.}\ \bibnamefont
  {Ter-Mikaelian}},\ }\href@noop {} {\emph {\bibinfo {title} {High-energy
  electromagnetic processes in condensed media}}},\ \bibinfo {number} {29}\
  (\bibinfo  {publisher} {John Wiley \& Sons},\ \bibinfo {year}
  {1972})\BibitemShut {NoStop}%
\bibitem [{\citenamefont {Lohmann}\ \emph {et~al.}(1994)\citenamefont
  {Lohmann}, \citenamefont {Peise}, \citenamefont {Ahrens}, \citenamefont
  {Anthony}, \citenamefont {Arends}, \citenamefont {Beck}, \citenamefont
  {Crawford}, \citenamefont {Hünger}, \citenamefont {Kaiser}, \citenamefont
  {Kellie} \emph {et~al.}}]{Lohmann1994Linearly}%
  \BibitemOpen
  \bibfield  {author} {\bibinfo {author} {\bibfnamefont {D.}~\bibnamefont
  {Lohmann}}, \bibinfo {author} {\bibfnamefont {J.}~\bibnamefont {Peise}},
  \bibinfo {author} {\bibfnamefont {J.}~\bibnamefont {Ahrens}}, \bibinfo
  {author} {\bibfnamefont {I.}~\bibnamefont {Anthony}}, \bibinfo {author}
  {\bibfnamefont {H.-J.}\ \bibnamefont {Arends}}, \bibinfo {author}
  {\bibfnamefont {R.}~\bibnamefont {Beck}}, \bibinfo {author} {\bibfnamefont
  {R.}~\bibnamefont {Crawford}}, \bibinfo {author} {\bibfnamefont
  {A.}~\bibnamefont {Hünger}}, \bibinfo {author} {\bibfnamefont {K.~H.}\
  \bibnamefont {Kaiser}}, \bibinfo {author} {\bibfnamefont {J.~D.}\
  \bibnamefont {Kellie}},  \emph {et~al.},\ }\bibfield  {title} {\enquote
  {\bibinfo {title} {{Linearly polarized photons at MAMI (Mainz)}},}\ }\href
  {\doibase https://doi.org/10.1016/0168-9002(94)90230-5} {\bibfield  {journal}
  {\bibinfo  {journal} {Nucl. Instrum. Methods Phys. Res., Sect. A}\ }\textbf
  {\bibinfo {volume} {343}},\ \bibinfo {pages} {494} (\bibinfo {year}
  {1994})}\BibitemShut {NoStop}%
\bibitem [{\citenamefont {Baier}\ \emph {et~al.}(1973)\citenamefont {Baier},
  \citenamefont {Katkov},\ and\ \citenamefont {Fadin}}]{baier1973radiation}%
  \BibitemOpen
  \bibfield  {author} {\bibinfo {author} {\bibfnamefont {V.~N.}\ \bibnamefont
  {Baier}}, \bibinfo {author} {\bibfnamefont {V.~M.}\ \bibnamefont {Katkov}}, \
  and\ \bibinfo {author} {\bibfnamefont {V.~S.}\ \bibnamefont {Fadin}},\
  }\href@noop {} {\emph {\bibinfo {title} {{Radiation of relativistic
  electrons; Izluchenie relyativistskikh elektronov}}}}\ (\bibinfo  {publisher}
  {Atomizdat, Moscow},\ \bibinfo {year} {1973})\BibitemShut {NoStop}%
\bibitem [{\citenamefont {Ritus}(1985)}]{ritus1985quantum}%
  \BibitemOpen
  \bibfield  {author} {\bibinfo {author} {\bibfnamefont {V.~I.}\ \bibnamefont
  {Ritus}},\ }\bibfield  {title} {\enquote {\bibinfo {title} {Quantum effects
  of the interaction of elementary particles with an intense electromagnetic
  field},}\ }\href {\doibase 10.1007/BF01120220} {\bibfield  {journal}
  {\bibinfo  {journal} {J. Sov. Laser Res.}\ }\textbf {\bibinfo {volume} {6}},\
  \bibinfo {pages} {497} (\bibinfo {year} {1985})}\BibitemShut {NoStop}%
\bibitem [{\citenamefont {Sokolov}\ and\ \citenamefont
  {Ternov}(1986)}]{sokolov1986radiation}%
  \BibitemOpen
  \bibfield  {author} {\bibinfo {author} {\bibfnamefont {A.~A.}\ \bibnamefont
  {Sokolov}}\ and\ \bibinfo {author} {\bibfnamefont {I.~M.}\ \bibnamefont
  {Ternov}},\ }\href@noop {} {\emph {\bibinfo {title} {Radiation from
  relativistic electrons}}}\ (\bibinfo  {publisher} {American Institute of
  Physics},\ \bibinfo {address} {New York},\ \bibinfo {year}
  {1986})\BibitemShut {NoStop}%
\bibitem [{\citenamefont {Khokonov}\ and\ \citenamefont
  {Bekulova}(2010)}]{khokonov2010length}%
  \BibitemOpen
  \bibfield  {author} {\bibinfo {author} {\bibfnamefont {M.~Kh.}\ \bibnamefont
  {Khokonov}}\ and\ \bibinfo {author} {\bibfnamefont {I.~Z.}\ \bibnamefont
  {Bekulova}},\ }\bibfield  {title} {\enquote {\bibinfo {title} {Length of
  formation of processes in a constant external field at high energies},}\
  }\href {\doibase 10.1134/S106378421005021X} {\bibfield  {journal} {\bibinfo
  {journal} {Tech. Phys.}\ }\textbf {\bibinfo {volume} {55}},\ \bibinfo {pages}
  {728} (\bibinfo {year} {2010})}\BibitemShut {NoStop}%
\bibitem [{\citenamefont {Omori}\ \emph {et~al.}(2006)\citenamefont {Omori},
  \citenamefont {Fukuda}, \citenamefont {Hirose}, \citenamefont {Kurihara},
  \citenamefont {Kuroda}, \citenamefont {Nomura}, \citenamefont {Ohashi},
  \citenamefont {Okugi}, \citenamefont {Sakaue}, \citenamefont {Saito} \emph
  {et~al.}}]{Omori2006efficient}%
  \BibitemOpen
  \bibfield  {author} {\bibinfo {author} {\bibfnamefont {T.}~\bibnamefont
  {Omori}}, \bibinfo {author} {\bibfnamefont {M.}~\bibnamefont {Fukuda}},
  \bibinfo {author} {\bibfnamefont {T.}~\bibnamefont {Hirose}}, \bibinfo
  {author} {\bibfnamefont {Y.}~\bibnamefont {Kurihara}}, \bibinfo {author}
  {\bibfnamefont {R.}~\bibnamefont {Kuroda}}, \bibinfo {author} {\bibfnamefont
  {M.}~\bibnamefont {Nomura}}, \bibinfo {author} {\bibfnamefont
  {A.}~\bibnamefont {Ohashi}}, \bibinfo {author} {\bibfnamefont
  {T.}~\bibnamefont {Okugi}}, \bibinfo {author} {\bibfnamefont
  {K.}~\bibnamefont {Sakaue}}, \bibinfo {author} {\bibfnamefont
  {T.}~\bibnamefont {Saito}},  \emph {et~al.},\ }\bibfield  {title} {\enquote
  {\bibinfo {title} {Efficient propagation of polarization from laser photons
  to positrons through {Compton} scattering and electron-positron pair
  creation},}\ }\href {\doibase 10.1103/PhysRevLett.96.114801} {\bibfield
  {journal} {\bibinfo  {journal} {Phys. Rev. Lett.}\ }\textbf {\bibinfo
  {volume} {96}},\ \bibinfo {pages} {114801} (\bibinfo {year}
  {2006})}\BibitemShut {NoStop}%
\bibitem [{\citenamefont {Alexander}\ \emph {et~al.}(2008)\citenamefont
  {Alexander}, \citenamefont {Barley}, \citenamefont {Batygin}, \citenamefont
  {Berridge}, \citenamefont {Bharadwaj}, \citenamefont {Bower}, \citenamefont
  {Bugg}, \citenamefont {Decker}, \citenamefont {Dollan}, \citenamefont
  {Efremenko} \emph {et~al.}}]{Alexander2008Observation}%
  \BibitemOpen
  \bibfield  {author} {\bibinfo {author} {\bibfnamefont {G.}~\bibnamefont
  {Alexander}}, \bibinfo {author} {\bibfnamefont {J.}~\bibnamefont {Barley}},
  \bibinfo {author} {\bibfnamefont {Y.}~\bibnamefont {Batygin}}, \bibinfo
  {author} {\bibfnamefont {S.}~\bibnamefont {Berridge}}, \bibinfo {author}
  {\bibfnamefont {V.}~\bibnamefont {Bharadwaj}}, \bibinfo {author}
  {\bibfnamefont {G.}~\bibnamefont {Bower}}, \bibinfo {author} {\bibfnamefont
  {W.}~\bibnamefont {Bugg}}, \bibinfo {author} {\bibfnamefont {F.-J.}\
  \bibnamefont {Decker}}, \bibinfo {author} {\bibfnamefont {R.}~\bibnamefont
  {Dollan}}, \bibinfo {author} {\bibfnamefont {Y.}~\bibnamefont {Efremenko}},
  \emph {et~al.},\ }\bibfield  {title} {\enquote {\bibinfo {title} {Observation
  of polarized positrons from an undulator-based source},}\ }\href {\doibase
  10.1103/PhysRevLett.100.210801} {\bibfield  {journal} {\bibinfo  {journal}
  {Phys. Rev. Lett.}\ }\textbf {\bibinfo {volume} {100}},\ \bibinfo {pages}
  {210801} (\bibinfo {year} {2008})}\BibitemShut {NoStop}%
\bibitem [{\citenamefont {Petrillo}\ \emph {et~al.}(2015)\citenamefont
  {Petrillo}, \citenamefont {Bacci}, \citenamefont {Curatolo}, \citenamefont
  {Drebot}, \citenamefont {Giribono}, \citenamefont {Maroli}, \citenamefont
  {Rossi}, \citenamefont {Serafini}, \citenamefont {Tomassini}, \citenamefont
  {Vaccarezza},\ and\ \citenamefont {Variola}}]{Petrillo2015Polarization}%
  \BibitemOpen
  \bibfield  {author} {\bibinfo {author} {\bibfnamefont {V.}~\bibnamefont
  {Petrillo}}, \bibinfo {author} {\bibfnamefont {A.}~\bibnamefont {Bacci}},
  \bibinfo {author} {\bibfnamefont {C.}~\bibnamefont {Curatolo}}, \bibinfo
  {author} {\bibfnamefont {I.}~\bibnamefont {Drebot}}, \bibinfo {author}
  {\bibfnamefont {A.}~\bibnamefont {Giribono}}, \bibinfo {author}
  {\bibfnamefont {C.}~\bibnamefont {Maroli}}, \bibinfo {author} {\bibfnamefont
  {A.~R.}\ \bibnamefont {Rossi}}, \bibinfo {author} {\bibfnamefont
  {L.}~\bibnamefont {Serafini}}, \bibinfo {author} {\bibfnamefont
  {P.}~\bibnamefont {Tomassini}}, \bibinfo {author} {\bibfnamefont
  {C.}~\bibnamefont {Vaccarezza}}, \ and\ \bibinfo {author} {\bibfnamefont
  {A.}~\bibnamefont {Variola}},\ }\bibfield  {title} {\enquote {\bibinfo
  {title} {Polarization of x-gamma radiation produced by a {Thomson} and
  {Compton} inverse scattering},}\ }\href {\doibase
  10.1103/PhysRevSTAB.18.110701} {\bibfield  {journal} {\bibinfo  {journal}
  {Phys. Rev. ST Accel. Beams}\ }\textbf {\bibinfo {volume} {18}},\ \bibinfo
  {pages} {110701} (\bibinfo {year} {2015})}\BibitemShut {NoStop}%
\bibitem [{\citenamefont {Wan}\ \emph {et~al.}(2020)\citenamefont {Wan},
  \citenamefont {Wang}, \citenamefont {Guo}, \citenamefont {Chen},
  \citenamefont {Shaisultanov}, \citenamefont {Xu}, \citenamefont
  {Hatsagortsyan}, \citenamefont {Keitel},\ and\ \citenamefont
  {Li}}]{Wan2020High}%
  \BibitemOpen
  \bibfield  {author} {\bibinfo {author} {\bibfnamefont {F.}~\bibnamefont
  {Wan}}, \bibinfo {author} {\bibfnamefont {Y.}~\bibnamefont {Wang}}, \bibinfo
  {author} {\bibfnamefont {R.~T.}\ \bibnamefont {Guo}}, \bibinfo {author}
  {\bibfnamefont {Y.~Y.}\ \bibnamefont {Chen}}, \bibinfo {author}
  {\bibfnamefont {R.}~\bibnamefont {Shaisultanov}}, \bibinfo {author}
  {\bibfnamefont {Z.~F.}\ \bibnamefont {Xu}}, \bibinfo {author} {\bibfnamefont
  {K.~Z.}\ \bibnamefont {Hatsagortsyan}}, \bibinfo {author} {\bibfnamefont
  {C.~H.}\ \bibnamefont {Keitel}}, \ and\ \bibinfo {author} {\bibfnamefont
  {J.~X.}\ \bibnamefont {Li}},\ }\bibfield  {title} {\enquote {\bibinfo {title}
  {{High-energy $\gamma$-photon polarization in nonlinear Breit-Wheeler pair
  production and $\ensuremath{\gamma}$ polarimetry}},}\ }\href {\doibase
  10.1103/PhysRevResearch.2.032049} {\bibfield  {journal} {\bibinfo  {journal}
  {Phys. Rev. Research}\ }\textbf {\bibinfo {volume} {2}},\ \bibinfo {pages}
  {032049} (\bibinfo {year} {2020})}\BibitemShut {NoStop}%
\bibitem [{\citenamefont {Li}\ \emph {et~al.}(2025{\natexlab{a}})\citenamefont
  {Li}, \citenamefont {Wang}, \citenamefont {Salamin}, \citenamefont
  {Ababekri}, \citenamefont {Wan}, \citenamefont {Zhao}, \citenamefont {Xue},
  \citenamefont {Tian},\ and\ \citenamefont {Li}}]{li2025generation}%
  \BibitemOpen
  \bibfield  {author} {\bibinfo {author} {\bibfnamefont {Z.~P.}\ \bibnamefont
  {Li}}, \bibinfo {author} {\bibfnamefont {Y.}~\bibnamefont {Wang}}, \bibinfo
  {author} {\bibfnamefont {Y.~I.}\ \bibnamefont {Salamin}}, \bibinfo {author}
  {\bibfnamefont {M.}~\bibnamefont {Ababekri}}, \bibinfo {author}
  {\bibfnamefont {F.}~\bibnamefont {Wan}}, \bibinfo {author} {\bibfnamefont
  {Q.}~\bibnamefont {Zhao}}, \bibinfo {author} {\bibfnamefont {K.}~\bibnamefont
  {Xue}}, \bibinfo {author} {\bibfnamefont {Y.}~\bibnamefont {Tian}}, \ and\
  \bibinfo {author} {\bibfnamefont {J.~X.}\ \bibnamefont {Li}},\ }\href
  {https://arxiv.org/abs/2504.11113} {\enquote {\bibinfo {title} {Generation of
  relativistic structured spin-polarized lepton beams},}\ } (\bibinfo {year}
  {2025}{\natexlab{a}}),\ \Eprint {http://arxiv.org/abs/2504.11113}
  {arXiv:2504.11113} \BibitemShut {NoStop}%
\bibitem [{\citenamefont {Li}\ \emph {et~al.}(2025{\natexlab{b}})\citenamefont
  {Li}, \citenamefont {Wang}, \citenamefont {Sun}, \citenamefont {Wan},
  \citenamefont {Salamin}, \citenamefont {Ababekri}, \citenamefont {Zhao},
  \citenamefont {Xue}, \citenamefont {Tian}, \citenamefont {Wei} \emph
  {et~al.}}]{li2025ultrafast}%
  \BibitemOpen
  \bibfield  {author} {\bibinfo {author} {\bibfnamefont {Z.~P.}\ \bibnamefont
  {Li}}, \bibinfo {author} {\bibfnamefont {Y.}~\bibnamefont {Wang}}, \bibinfo
  {author} {\bibfnamefont {T.}~\bibnamefont {Sun}}, \bibinfo {author}
  {\bibfnamefont {F.}~\bibnamefont {Wan}}, \bibinfo {author} {\bibfnamefont
  {Y.~I.}\ \bibnamefont {Salamin}}, \bibinfo {author} {\bibfnamefont
  {M.}~\bibnamefont {Ababekri}}, \bibinfo {author} {\bibfnamefont
  {Q.}~\bibnamefont {Zhao}}, \bibinfo {author} {\bibfnamefont {K.}~\bibnamefont
  {Xue}}, \bibinfo {author} {\bibfnamefont {Y.}~\bibnamefont {Tian}}, \bibinfo
  {author} {\bibfnamefont {W.~Q.}\ \bibnamefont {Wei}},  \emph {et~al.},\
  }\bibfield  {title} {\enquote {\bibinfo {title} {Ultrafast spin rotation of
  relativistic lepton beams via terahertz wave in a dielectric-lined
  waveguide},}\ }\href {\doibase 10.1103/PhysRevLett.134.075001} {\bibfield
  {journal} {\bibinfo  {journal} {Phys. Rev. Lett.}\ }\textbf {\bibinfo
  {volume} {134}},\ \bibinfo {pages} {075001} (\bibinfo {year}
  {2025}{\natexlab{b}})}\BibitemShut {NoStop}%
\bibitem [{SM()}]{SM}%
  \BibitemOpen
  \href@noop {} {}\bibinfo {note} {See Supplemental Materials for details on
  the analysis of photon polarization in radially polarized fields, the role of
  continuous focusing in suppressing polarization cancellation, the influence
  of other beam and target parameters, the effects of Bremsstrahlung and
  Bethe-Heitler processes, the polarization of the seed electron beam, and
  other noteworthy configurations.}\BibitemShut {Stop}%
\bibitem [{\citenamefont {Anselmino}\ \emph {et~al.}(1995)\citenamefont
  {Anselmino}, \citenamefont {Efremov},\ and\ \citenamefont
  {Leader}}]{Anselmino1995The}%
  \BibitemOpen
  \bibfield  {author} {\bibinfo {author} {\bibfnamefont {M.}~\bibnamefont
  {Anselmino}}, \bibinfo {author} {\bibfnamefont {A.}~\bibnamefont {Efremov}},
  \ and\ \bibinfo {author} {\bibfnamefont {E.}~\bibnamefont {Leader}},\
  }\bibfield  {title} {\enquote {\bibinfo {title} {The theory and phenomenology
  of polarized deep-inelastic scattering},}\ }\href {\doibase
  10.1016/0370-1573(95)00011-5} {\bibfield  {journal} {\bibinfo  {journal}
  {Phys. Rep.}\ }\textbf {\bibinfo {volume} {261}},\ \bibinfo {pages} {1}
  (\bibinfo {year} {1995})}\BibitemShut {NoStop}%
\bibitem [{\citenamefont {Hughes}\ and\ \citenamefont
  {Voss}(1999)}]{Hughes1999Spin}%
  \BibitemOpen
  \bibfield  {author} {\bibinfo {author} {\bibfnamefont {E.~W.}\ \bibnamefont
  {Hughes}}\ and\ \bibinfo {author} {\bibfnamefont {R.}~\bibnamefont {Voss}},\
  }\bibfield  {title} {\enquote {\bibinfo {title} {Spin structure functions},}\
  }\href {\doibase 10.1146/annurev.nucl.49.1.303} {\bibfield  {journal}
  {\bibinfo  {journal} {Annu. Rev. Nucl. Part. Sci.}\ }\textbf {\bibinfo
  {volume} {49}},\ \bibinfo {pages} {303} (\bibinfo {year} {1999})},\ \bibinfo
  {note} {86}\BibitemShut {NoStop}%
\bibitem [{\citenamefont {Bluemlein}(2013)}]{Bluemlein2013The}%
  \BibitemOpen
  \bibfield  {author} {\bibinfo {author} {\bibfnamefont {J.}~\bibnamefont
  {Bluemlein}},\ }\bibfield  {title} {\enquote {\bibinfo {title} {The theory of
  deeply inelastic scattering},}\ }\href {\doibase 10.1016/j.ppnp.2012.09.006}
  {\bibfield  {journal} {\bibinfo  {journal} {Prog. Part. Nucl. Phys.}\
  }\textbf {\bibinfo {volume} {69}},\ \bibinfo {pages} {28} (\bibinfo {year}
  {2013})}\BibitemShut {NoStop}%
\bibitem [{\citenamefont {Abelev}\ \emph {et~al.}(2009)\citenamefont {Abelev},
  \citenamefont {Aggarwal}, \citenamefont {Ahammed}, \citenamefont
  {Alakhverdyants}, \citenamefont {Anderson}, \citenamefont {Arkhipkin},
  \citenamefont {Averichev}, \citenamefont {Balewski}, \citenamefont
  {Barannikova}, \citenamefont {Barnby} \emph
  {et~al.}}]{Abelev2009Longitudinal}%
  \BibitemOpen
  \bibfield  {author} {\bibinfo {author} {\bibfnamefont {B.~I.}\ \bibnamefont
  {Abelev}}, \bibinfo {author} {\bibfnamefont {M.~M.}\ \bibnamefont
  {Aggarwal}}, \bibinfo {author} {\bibfnamefont {Z.}~\bibnamefont {Ahammed}},
  \bibinfo {author} {\bibfnamefont {A.~V.}\ \bibnamefont {Alakhverdyants}},
  \bibinfo {author} {\bibfnamefont {B.~D.}\ \bibnamefont {Anderson}}, \bibinfo
  {author} {\bibfnamefont {D.}~\bibnamefont {Arkhipkin}}, \bibinfo {author}
  {\bibfnamefont {G.~S.}\ \bibnamefont {Averichev}}, \bibinfo {author}
  {\bibfnamefont {J.}~\bibnamefont {Balewski}}, \bibinfo {author}
  {\bibfnamefont {O.}~\bibnamefont {Barannikova}}, \bibinfo {author}
  {\bibfnamefont {L.~S.}\ \bibnamefont {Barnby}},  \emph {et~al.} (\bibinfo
  {collaboration} {STAR Collaboration}),\ }\bibfield  {title} {\enquote
  {\bibinfo {title} {Longitudinal spin transfer to $\ensuremath{\Lambda}$ and
  $\overline{\ensuremath{\Lambda}}$ hyperons in polarized proton-proton
  collisions at $\sqrt{s}=200\text{ }\text{ }\mathrm{GeV}$},}\ }\href {\doibase
  10.1103/PhysRevD.80.111102} {\bibfield  {journal} {\bibinfo  {journal} {Phys.
  Rev. D}\ }\textbf {\bibinfo {volume} {80}},\ \bibinfo {pages} {111102}
  (\bibinfo {year} {2009})}\BibitemShut {NoStop}%
\bibitem [{\citenamefont {Adamczyk}\ \emph {et~al.}(2014)\citenamefont
  {Adamczyk}, \citenamefont {Adkins}, \citenamefont {Agakishiev}, \citenamefont
  {Aggarwal}, \citenamefont {Ahammed}, \citenamefont {Alekseev}, \citenamefont
  {Alford}, \citenamefont {Anson}, \citenamefont {Aparin}, \citenamefont
  {Arkhipkin} \emph {et~al.}}]{Adamczyk2014Measurement}%
  \BibitemOpen
  \bibfield  {author} {\bibinfo {author} {\bibfnamefont {L.}~\bibnamefont
  {Adamczyk}}, \bibinfo {author} {\bibfnamefont {J.~K.}\ \bibnamefont
  {Adkins}}, \bibinfo {author} {\bibfnamefont {G.}~\bibnamefont {Agakishiev}},
  \bibinfo {author} {\bibfnamefont {M.~M.}\ \bibnamefont {Aggarwal}}, \bibinfo
  {author} {\bibfnamefont {Z.}~\bibnamefont {Ahammed}}, \bibinfo {author}
  {\bibfnamefont {I.}~\bibnamefont {Alekseev}}, \bibinfo {author}
  {\bibfnamefont {J.}~\bibnamefont {Alford}}, \bibinfo {author} {\bibfnamefont
  {C.~D.}\ \bibnamefont {Anson}}, \bibinfo {author} {\bibfnamefont
  {A.}~\bibnamefont {Aparin}}, \bibinfo {author} {\bibfnamefont
  {D.}~\bibnamefont {Arkhipkin}},  \emph {et~al.} (\bibinfo {collaboration}
  {STAR Collaboration}),\ }\bibfield  {title} {\enquote {\bibinfo {title}
  {Measurement of longitudinal spin asymmetries for weak boson production in
  polarized proton-proton collisions at {RHIC}},}\ }\href {\doibase
  10.1103/PhysRevLett.113.072301} {\bibfield  {journal} {\bibinfo  {journal}
  {Phys. Rev. Lett.}\ }\textbf {\bibinfo {volume} {113}},\ \bibinfo {pages}
  {072301} (\bibinfo {year} {2014})}\BibitemShut {NoStop}%
\bibitem [{\citenamefont {Rosenberg}(2011)}]{Rosenberg2011Spin}%
  \BibitemOpen
  \bibfield  {author} {\bibinfo {author} {\bibfnamefont {R.~A.}\ \bibnamefont
  {Rosenberg}},\ }\enquote {\bibinfo {title} {Spin-polarized electron induced
  asymmetric reactions in chiral molecules},}\ in\ \href {\doibase
  10.1007/128_2010_81} {\emph {\bibinfo {booktitle} {Electronic and Magnetic
  Properties of Chiral Molecules and Supramolecular Architectures}}},\ \bibinfo
  {series} {Topics in Current Chemistry-Series}, Vol.\ \bibinfo {volume}
  {298},\ \bibinfo {editor} {edited by\ \bibinfo {editor} {\bibfnamefont
  {R.}~\bibnamefont {Naaman}}, \bibinfo {editor} {\bibfnamefont {D.~N.}\
  \bibnamefont {Beratan}}, \ and\ \bibinfo {editor} {\bibfnamefont {D.~H.}\
  \bibnamefont {Waldeck}}}\ (\bibinfo  {publisher} {Springer Berlin
  Heidelberg},\ \bibinfo {year} {2011})\ p.\ \bibinfo {pages} {279}\BibitemShut
  {NoStop}%
\bibitem [{\citenamefont {Bonner}(1991)}]{Bonner1991The}%
  \BibitemOpen
  \bibfield  {author} {\bibinfo {author} {\bibfnamefont {W.~A.}\ \bibnamefont
  {Bonner}},\ }\bibfield  {title} {\enquote {\bibinfo {title} {The origin and
  amplification of biomolecular chirality},}\ }\href {\doibase
  10.1007/bf01809580} {\bibfield  {journal} {\bibinfo  {journal} {Orig. Life
  Evol. Biosph.}\ }\textbf {\bibinfo {volume} {21}},\ \bibinfo {pages} {59}
  (\bibinfo {year} {1991})},\ \bibinfo {note} {467}\BibitemShut {NoStop}%
\bibitem [{\citenamefont {Kessler}(1985)}]{kessler85polarized}%
  \BibitemOpen
  \bibfield  {author} {\bibinfo {author} {\bibfnamefont {J.}~\bibnamefont
  {Kessler}},\ }\href@noop {} {\emph {\bibinfo {title} {Polarized
  Electrons}}},\ Vol.~\bibinfo {volume} {1}\ (\bibinfo  {publisher} {Springer
  Science \& Business Media},\ \bibinfo {year} {1985})\BibitemShut {NoStop}%
\bibitem [{\citenamefont {Moortgat-Pick}\ \emph {et~al.}(2008)\citenamefont
  {Moortgat-Pick}, \citenamefont {Abe}, \citenamefont {Alexander},
  \citenamefont {Ananthanarayan}, \citenamefont {Babich}, \citenamefont
  {Bharadwaj}, \citenamefont {Barber}, \citenamefont {Bartl}, \citenamefont
  {Brachmann}, \citenamefont {Chen} \emph {et~al.}}]{moortgat2008polarized}%
  \BibitemOpen
  \bibfield  {author} {\bibinfo {author} {\bibfnamefont {G.}~\bibnamefont
  {Moortgat-Pick}}, \bibinfo {author} {\bibfnamefont {T.}~\bibnamefont {Abe}},
  \bibinfo {author} {\bibfnamefont {G.}~\bibnamefont {Alexander}}, \bibinfo
  {author} {\bibfnamefont {B.}~\bibnamefont {Ananthanarayan}}, \bibinfo
  {author} {\bibfnamefont {A.~A.}\ \bibnamefont {Babich}}, \bibinfo {author}
  {\bibfnamefont {V.}~\bibnamefont {Bharadwaj}}, \bibinfo {author}
  {\bibfnamefont {D.}~\bibnamefont {Barber}}, \bibinfo {author} {\bibfnamefont
  {A.}~\bibnamefont {Bartl}}, \bibinfo {author} {\bibfnamefont
  {A.}~\bibnamefont {Brachmann}}, \bibinfo {author} {\bibfnamefont
  {Si}~\bibnamefont {Chen}},  \emph {et~al.},\ }\bibfield  {title} {\enquote
  {\bibinfo {title} {Polarized positrons and electrons at the linear
  collider},}\ }\href {\doibase https://doi.org/10.1016/j.physrep.2007.12.003}
  {\bibfield  {journal} {\bibinfo  {journal} {Phys. Rep.}\ }\textbf {\bibinfo
  {volume} {460}},\ \bibinfo {pages} {131} (\bibinfo {year}
  {2008})}\BibitemShut {NoStop}%
\bibitem [{\citenamefont {Akbar}\ \emph {et~al.}(2017)\citenamefont {Akbar},
  \citenamefont {Roy}, \citenamefont {Park}, \citenamefont {Crede},
  \citenamefont {Anisovich}, \citenamefont {Denisenko}, \citenamefont {Klempt},
  \citenamefont {Nikonov}, \citenamefont {Sarantsev}, \citenamefont {Adhikari}
  \emph {et~al.}}]{Akbar2017Measurement}%
  \BibitemOpen
  \bibfield  {author} {\bibinfo {author} {\bibfnamefont {Z.}~\bibnamefont
  {Akbar}}, \bibinfo {author} {\bibfnamefont {P.}~\bibnamefont {Roy}}, \bibinfo
  {author} {\bibfnamefont {S.}~\bibnamefont {Park}}, \bibinfo {author}
  {\bibfnamefont {V.}~\bibnamefont {Crede}}, \bibinfo {author} {\bibfnamefont
  {A.~V.}\ \bibnamefont {Anisovich}}, \bibinfo {author} {\bibfnamefont
  {I.}~\bibnamefont {Denisenko}}, \bibinfo {author} {\bibfnamefont
  {E.}~\bibnamefont {Klempt}}, \bibinfo {author} {\bibfnamefont {V.~A.}\
  \bibnamefont {Nikonov}}, \bibinfo {author} {\bibfnamefont {A.~V.}\
  \bibnamefont {Sarantsev}}, \bibinfo {author} {\bibfnamefont {K.~P.}\
  \bibnamefont {Adhikari}},  \emph {et~al.} (\bibinfo {collaboration} {The CLAS
  Collaboration}),\ }\bibfield  {title} {\enquote {\bibinfo {title}
  {Measurement of the helicity asymmetry $e$ in
  $\ensuremath{\omega}\ensuremath{\rightarrow}{\ensuremath{\pi}}^{+}{\ensuremath{\pi}}^{\ensuremath{-}}{\ensuremath{\pi}}^{0}$
  photoproduction},}\ }\href {\doibase 10.1103/PhysRevC.96.065209} {\bibfield
  {journal} {\bibinfo  {journal} {Phys. Rev. C}\ }\textbf {\bibinfo {volume}
  {96}},\ \bibinfo {pages} {065209} (\bibinfo {year} {2017})}\BibitemShut
  {NoStop}%
\bibitem [{\citenamefont {Bragin}\ \emph {et~al.}(2017)\citenamefont {Bragin},
  \citenamefont {Meuren}, \citenamefont {Keitel},\ and\ \citenamefont
  {Di~Piazza}}]{bragin2017high}%
  \BibitemOpen
  \bibfield  {author} {\bibinfo {author} {\bibfnamefont {S.}~\bibnamefont
  {Bragin}}, \bibinfo {author} {\bibfnamefont {S.}~\bibnamefont {Meuren}},
  \bibinfo {author} {\bibfnamefont {C.~H.}\ \bibnamefont {Keitel}}, \ and\
  \bibinfo {author} {\bibfnamefont {A.}~\bibnamefont {Di~Piazza}},\ }\bibfield
  {title} {\enquote {\bibinfo {title} {High-energy vacuum birefringence and
  dichroism in an ultrastrong laser field},}\ }\href {\doibase
  10.1103/PhysRevLett.119.250403} {\bibfield  {journal} {\bibinfo  {journal}
  {Phys. Rev. Lett.}\ }\textbf {\bibinfo {volume} {119}},\ \bibinfo {pages}
  {250403} (\bibinfo {year} {2017})}\BibitemShut {NoStop}%
\bibitem [{\citenamefont {Wan}\ \emph {et~al.}(2023)\citenamefont {Wan},
  \citenamefont {Lv}, \citenamefont {Xue}, \citenamefont {Dou}, \citenamefont
  {Zhao}, \citenamefont {Ababekri}, \citenamefont {Wei}, \citenamefont {Li},
  \citenamefont {Zhao},\ and\ \citenamefont {Li}}]{wan2023simulations}%
  \BibitemOpen
  \bibfield  {author} {\bibinfo {author} {\bibfnamefont {F.}~\bibnamefont
  {Wan}}, \bibinfo {author} {\bibfnamefont {C.}~\bibnamefont {Lv}}, \bibinfo
  {author} {\bibfnamefont {K.}~\bibnamefont {Xue}}, \bibinfo {author}
  {\bibfnamefont {Z.-K.}\ \bibnamefont {Dou}}, \bibinfo {author} {\bibfnamefont
  {Q.}~\bibnamefont {Zhao}}, \bibinfo {author} {\bibfnamefont {M.}~\bibnamefont
  {Ababekri}}, \bibinfo {author} {\bibfnamefont {W.-Q.}\ \bibnamefont {Wei}},
  \bibinfo {author} {\bibfnamefont {Z.-P.}\ \bibnamefont {Li}}, \bibinfo
  {author} {\bibfnamefont {Y.-T.}\ \bibnamefont {Zhao}}, \ and\ \bibinfo
  {author} {\bibfnamefont {J.-X.}\ \bibnamefont {Li}},\ }\bibfield  {title}
  {\enquote {\bibinfo {title} {Simulations of spin/polarization-resolved
  laser–plasma interactions in the nonlinear {QED} regime},}\ }\href
  {\doibase 10.1063/5.0163929} {\bibfield  {journal} {\bibinfo  {journal}
  {Matter Radiat. Extremes}\ }\textbf {\bibinfo {volume} {8}},\ \bibinfo
  {pages} {064002} (\bibinfo {year} {2023})}\BibitemShut {NoStop}%
\bibitem [{\citenamefont {Baier}\ \emph {et~al.}(1998)\citenamefont {Baier},
  \citenamefont {Katkov},\ and\ \citenamefont
  {Strakhovenko}}]{baier1998Electromagnetic}%
  \BibitemOpen
  \bibfield  {author} {\bibinfo {author} {\bibfnamefont {V.~N.}\ \bibnamefont
  {Baier}}, \bibinfo {author} {\bibfnamefont {V.~M.}\ \bibnamefont {Katkov}}, \
  and\ \bibinfo {author} {\bibfnamefont {V.~M.}\ \bibnamefont {Strakhovenko}},\
  }\href@noop {} {\emph {\bibinfo {title} {{Electromagnetic Processes at High
  Energies in Oriented Single Crystals}}}}\ (\bibinfo  {publisher} {World
  Scientific},\ \bibinfo {address} {Singapore},\ \bibinfo {year}
  {1998})\BibitemShut {NoStop}%
\bibitem [{\citenamefont {Ammosov}\ \emph {et~al.}(1986)\citenamefont
  {Ammosov}, \citenamefont {Delone},\ and\ \citenamefont
  {Krainov}}]{Ammosov1986tunnel}%
  \BibitemOpen
  \bibfield  {author} {\bibinfo {author} {\bibfnamefont {M.~V.}\ \bibnamefont
  {Ammosov}}, \bibinfo {author} {\bibfnamefont {N.~B.}\ \bibnamefont {Delone}},
  \ and\ \bibinfo {author} {\bibfnamefont {V.~P.}\ \bibnamefont {Krainov}},\
  }\bibfield  {title} {\enquote {\bibinfo {title} {Tunnel ionization of complex
  atoms and of atomic ions in an alternating electromagnetic field},}\
  }\href@noop {} {\bibfield  {journal} {\bibinfo  {journal} {Sov. Phys. JETP}\
  }\textbf {\bibinfo {volume} {64}},\ \bibinfo {pages} {1191} (\bibinfo {year}
  {1986})}\BibitemShut {NoStop}%
\bibitem [{\citenamefont {Posthumus}\ \emph {et~al.}(1997)\citenamefont
  {Posthumus}, \citenamefont {Thompson}, \citenamefont {Frasinski},\ and\
  \citenamefont {Codling}}]{posthumus1997molecular}%
  \BibitemOpen
  \bibfield  {author} {\bibinfo {author} {\bibfnamefont {J.~H.}\ \bibnamefont
  {Posthumus}}, \bibinfo {author} {\bibfnamefont {M.~R.}\ \bibnamefont
  {Thompson}}, \bibinfo {author} {\bibfnamefont {L.~J.}\ \bibnamefont
  {Frasinski}}, \ and\ \bibinfo {author} {\bibfnamefont {K.}~\bibnamefont
  {Codling}},\ }\bibfield  {title} {\enquote {\bibinfo {title} {Molecular
  dissociative ionisation using a classical over-the-barrier approach},}\
  }\href@noop {} {\bibfield  {journal} {\bibinfo  {journal} {Multiphoton
  Processes 1996}\ }\textbf {\bibinfo {volume} {154}},\ \bibinfo {pages} {298}
  (\bibinfo {year} {1997})}\BibitemShut {NoStop}%
\bibitem [{\citenamefont {Yakimenko}\ \emph
  {et~al.}(2019{\natexlab{a}})\citenamefont {Yakimenko}, \citenamefont
  {Alsberg}, \citenamefont {Bong}, \citenamefont {Bouchard}, \citenamefont
  {Clarke}, \citenamefont {Emma}, \citenamefont {Green}, \citenamefont {Hast},
  \citenamefont {Hogan}, \citenamefont {Seabury} \emph
  {et~al.}}]{yakimenko2019FACET}%
  \BibitemOpen
  \bibfield  {author} {\bibinfo {author} {\bibfnamefont {V.}~\bibnamefont
  {Yakimenko}}, \bibinfo {author} {\bibfnamefont {L.}~\bibnamefont {Alsberg}},
  \bibinfo {author} {\bibfnamefont {E.}~\bibnamefont {Bong}}, \bibinfo {author}
  {\bibfnamefont {G.}~\bibnamefont {Bouchard}}, \bibinfo {author}
  {\bibfnamefont {C.}~\bibnamefont {Clarke}}, \bibinfo {author} {\bibfnamefont
  {C.}~\bibnamefont {Emma}}, \bibinfo {author} {\bibfnamefont {S.}~\bibnamefont
  {Green}}, \bibinfo {author} {\bibfnamefont {C.}~\bibnamefont {Hast}},
  \bibinfo {author} {\bibfnamefont {M.~J.}\ \bibnamefont {Hogan}}, \bibinfo
  {author} {\bibfnamefont {J.}~\bibnamefont {Seabury}},  \emph {et~al.},\
  }\bibfield  {title} {\enquote {\bibinfo {title} {{FACET-II} facility for
  advanced accelerator experimental tests},}\ }\href {\doibase
  10.1103/PhysRevAccelBeams.22.101301} {\bibfield  {journal} {\bibinfo
  {journal} {Phys. Rev. Accel. Beams}\ }\textbf {\bibinfo {volume} {22}},\
  \bibinfo {pages} {101301} (\bibinfo {year} {2019}{\natexlab{a}})}\BibitemShut
  {NoStop}%
\bibitem [{\citenamefont {Yakimenko}\ \emph
  {et~al.}(2019{\natexlab{b}})\citenamefont {Yakimenko}, \citenamefont
  {Meuren}, \citenamefont {Del~Gaudio}, \citenamefont {Baumann}, \citenamefont
  {Fedotov}, \citenamefont {Fiuza}, \citenamefont {Grismayer}, \citenamefont
  {Hogan}, \citenamefont {Pukhov}, \citenamefont {Silva},\ and\ \citenamefont
  {White}}]{yakimenko2019prospect}%
  \BibitemOpen
  \bibfield  {author} {\bibinfo {author} {\bibfnamefont {V.}~\bibnamefont
  {Yakimenko}}, \bibinfo {author} {\bibfnamefont {S.}~\bibnamefont {Meuren}},
  \bibinfo {author} {\bibfnamefont {F.}~\bibnamefont {Del~Gaudio}}, \bibinfo
  {author} {\bibfnamefont {C.}~\bibnamefont {Baumann}}, \bibinfo {author}
  {\bibfnamefont {A.}~\bibnamefont {Fedotov}}, \bibinfo {author} {\bibfnamefont
  {F.}~\bibnamefont {Fiuza}}, \bibinfo {author} {\bibfnamefont
  {T.}~\bibnamefont {Grismayer}}, \bibinfo {author} {\bibfnamefont {M.~J.}\
  \bibnamefont {Hogan}}, \bibinfo {author} {\bibfnamefont {A.}~\bibnamefont
  {Pukhov}}, \bibinfo {author} {\bibfnamefont {L.~O.}\ \bibnamefont {Silva}}, \
  and\ \bibinfo {author} {\bibfnamefont {G.}~\bibnamefont {White}},\ }\bibfield
   {title} {\enquote {\bibinfo {title} {Prospect of studying nonperturbative
  {QED} with beam-beam collisions},}\ }\href {\doibase
  10.1103/PhysRevLett.122.190404} {\bibfield  {journal} {\bibinfo  {journal}
  {Phys. Rev. Lett.}\ }\textbf {\bibinfo {volume} {122}},\ \bibinfo {pages}
  {190404} (\bibinfo {year} {2019}{\natexlab{b}})}\BibitemShut {NoStop}%
\bibitem [{\citenamefont {Emma}\ \emph {et~al.}(2025)\citenamefont {Emma},
  \citenamefont {Majernik}, \citenamefont {Swanson}, \citenamefont {Ariniello},
  \citenamefont {Gessner}, \citenamefont {Hessami}, \citenamefont {Hogan},
  \citenamefont {Knetsch}, \citenamefont {Larsen}, \citenamefont {Marinelli}
  \emph {et~al.}}]{Emma2025Experimental}%
  \BibitemOpen
  \bibfield  {author} {\bibinfo {author} {\bibfnamefont {C.}~\bibnamefont
  {Emma}}, \bibinfo {author} {\bibfnamefont {N.}~\bibnamefont {Majernik}},
  \bibinfo {author} {\bibfnamefont {K.~K.}\ \bibnamefont {Swanson}}, \bibinfo
  {author} {\bibfnamefont {R.}~\bibnamefont {Ariniello}}, \bibinfo {author}
  {\bibfnamefont {S.}~\bibnamefont {Gessner}}, \bibinfo {author} {\bibfnamefont
  {R.}~\bibnamefont {Hessami}}, \bibinfo {author} {\bibfnamefont {M.~J.}\
  \bibnamefont {Hogan}}, \bibinfo {author} {\bibfnamefont {A.}~\bibnamefont
  {Knetsch}}, \bibinfo {author} {\bibfnamefont {K.~A.}\ \bibnamefont {Larsen}},
  \bibinfo {author} {\bibfnamefont {A.}~\bibnamefont {Marinelli}},  \emph
  {et~al.},\ }\bibfield  {title} {\enquote {\bibinfo {title} {Experimental
  generation of extreme electron beams for advanced accelerator
  applications},}\ }\href {\doibase 10.1103/PhysRevLett.134.085001} {\bibfield
  {journal} {\bibinfo  {journal} {Phys. Rev. Lett.}\ }\textbf {\bibinfo
  {volume} {134}},\ \bibinfo {pages} {085001} (\bibinfo {year}
  {2025})}\BibitemShut {NoStop}%
\bibitem [{\citenamefont {Clarke}\ \emph {et~al.}(2022)\citenamefont {Clarke},
  \citenamefont {Esarey}, \citenamefont {Geddes}, \citenamefont {Hofstaetter},
  \citenamefont {Hogan}, \citenamefont {Nagaitsev}, \citenamefont {Palmer},
  \citenamefont {Piot}, \citenamefont {Power}, \citenamefont {Schroeder} \emph
  {et~al.}}]{Clarke2022advanced}%
  \BibitemOpen
  \bibfield  {author} {\bibinfo {author} {\bibfnamefont {C.}~\bibnamefont
  {Clarke}}, \bibinfo {author} {\bibfnamefont {E.}~\bibnamefont {Esarey}},
  \bibinfo {author} {\bibfnamefont {C.}~\bibnamefont {Geddes}}, \bibinfo
  {author} {\bibfnamefont {G.}~\bibnamefont {Hofstaetter}}, \bibinfo {author}
  {\bibfnamefont {M.J.}\ \bibnamefont {Hogan}}, \bibinfo {author}
  {\bibfnamefont {S.}~\bibnamefont {Nagaitsev}}, \bibinfo {author}
  {\bibfnamefont {M.}~\bibnamefont {Palmer}}, \bibinfo {author} {\bibfnamefont
  {P.}~\bibnamefont {Piot}}, \bibinfo {author} {\bibfnamefont {J.}~\bibnamefont
  {Power}}, \bibinfo {author} {\bibfnamefont {C.}~\bibnamefont {Schroeder}},
  \emph {et~al.},\ }\bibfield  {title} {\enquote {\bibinfo {title}
  {{U.S.} advanced and novel accelerator beam test facilities},}\ }\href
  {\doibase 10.1088/1748-0221/17/05/T05009} {\bibfield  {journal} {\bibinfo
  {journal} {J. Instrum.}\ }\textbf {\bibinfo {volume} {17}},\ \bibinfo {pages}
  {T05009} (\bibinfo {year} {2022})}\BibitemShut {NoStop}%
\bibitem [{\citenamefont {Babjak}\ \emph {et~al.}(2024)\citenamefont {Babjak},
  \citenamefont {Willingale}, \citenamefont {Arefiev},\ and\ \citenamefont
  {Vranic}}]{babjak2024direct}%
  \BibitemOpen
  \bibfield  {author} {\bibinfo {author} {\bibfnamefont {R.}~\bibnamefont
  {Babjak}}, \bibinfo {author} {\bibfnamefont {L.}~\bibnamefont {Willingale}},
  \bibinfo {author} {\bibfnamefont {A.}~\bibnamefont {Arefiev}}, \ and\
  \bibinfo {author} {\bibfnamefont {M.}~\bibnamefont {Vranic}},\ }\bibfield
  {title} {\enquote {\bibinfo {title} {Direct laser acceleration in underdense
  plasmas with multi-{PW} lasers: A path to high-charge, {GeV}-class electron
  bunches},}\ }\href {\doibase 10.1103/PhysRevLett.132.125001} {\bibfield
  {journal} {\bibinfo  {journal} {Phys. Rev. Lett.}\ }\textbf {\bibinfo
  {volume} {132}},\ \bibinfo {pages} {125001} (\bibinfo {year}
  {2024})}\BibitemShut {NoStop}%
\bibitem [{\citenamefont {Pukhov}\ \emph {et~al.}(1999)\citenamefont {Pukhov},
  \citenamefont {Sheng},\ and\ \citenamefont {Meyer-ter
  Vehn}}]{Pukhov1999Particle}%
  \BibitemOpen
  \bibfield  {author} {\bibinfo {author} {\bibfnamefont {A.}~\bibnamefont
  {Pukhov}}, \bibinfo {author} {\bibfnamefont {Z.-M.}\ \bibnamefont {Sheng}}, \
  and\ \bibinfo {author} {\bibfnamefont {J.}~\bibnamefont {Meyer-ter Vehn}},\
  }\bibfield  {title} {\enquote {\bibinfo {title} {Particle acceleration in
  relativistic laser channels},}\ }\href {\doibase 10.1063/1.873242} {\bibfield
   {journal} {\bibinfo  {journal} {Phys. Plasmas}\ }\textbf {\bibinfo {volume}
  {6}},\ \bibinfo {pages} {2847} (\bibinfo {year} {1999})},\ \Eprint
  {http://arxiv.org/abs/https://pubs.aip.org/aip/pop/article-pdf/6/7/2847/19054772/2847\_1\_online.pdf}
  {https://pubs.aip.org/aip/pop/article-pdf/6/7/2847/19054772/2847\_1\_online.pdf}
  \BibitemShut {NoStop}%
\bibitem [{\citenamefont {Aniculaesei}\ \emph {et~al.}(2023)\citenamefont
  {Aniculaesei}, \citenamefont {Ha}, \citenamefont {Yoffe}, \citenamefont
  {Labun}, \citenamefont {Milton}, \citenamefont {McCary}, \citenamefont
  {Spinks}, \citenamefont {Quevedo}, \citenamefont {Labun}, \citenamefont
  {Sain}, \citenamefont {Hannasch} \emph
  {et~al.}}]{aniculaesei2023acceleration}%
  \BibitemOpen
  \bibfield  {author} {\bibinfo {author} {\bibfnamefont {C.}~\bibnamefont
  {Aniculaesei}}, \bibinfo {author} {\bibfnamefont {T.}~\bibnamefont {Ha}},
  \bibinfo {author} {\bibfnamefont {S.}~\bibnamefont {Yoffe}}, \bibinfo
  {author} {\bibfnamefont {L.}~\bibnamefont {Labun}}, \bibinfo {author}
  {\bibfnamefont {S.}~\bibnamefont {Milton}}, \bibinfo {author} {\bibfnamefont
  {E.}~\bibnamefont {McCary}}, \bibinfo {author} {\bibfnamefont {M.~M.}\
  \bibnamefont {Spinks}}, \bibinfo {author} {\bibfnamefont {H.~J.}\
  \bibnamefont {Quevedo}}, \bibinfo {author} {\bibfnamefont {Ou~Z.}\
  \bibnamefont {Labun}}, \bibinfo {author} {\bibfnamefont {R.}~\bibnamefont
  {Sain}}, \bibinfo {author} {\bibfnamefont {A.}~\bibnamefont {Hannasch}},
  \emph {et~al.},\ }\bibfield  {title} {\enquote {\bibinfo {title} {The
  acceleration of a high-charge electron bunch to 10 {GeV} in a 10-cm
  nanoparticle-assisted wakefield accelerator},}\ }\href {\doibase
  10.1063/5.0161687} {\bibfield  {journal} {\bibinfo  {journal} {Matter Radiat.
  Extremes}\ }\textbf {\bibinfo {volume} {9}},\ \bibinfo {pages} {014001}
  (\bibinfo {year} {2023})}\BibitemShut {NoStop}%
\bibitem [{\citenamefont {Doss}\ \emph {et~al.}(2019)\citenamefont {Doss},
  \citenamefont {Adli}, \citenamefont {Ariniello}, \citenamefont {Cary},
  \citenamefont {Corde}, \citenamefont {Hidding}, \citenamefont {Hogan},
  \citenamefont {Hunt-Stone}, \citenamefont {Joshi}, \citenamefont {Marsh}
  \emph {et~al.}}]{doss2019laser}%
  \BibitemOpen
  \bibfield  {author} {\bibinfo {author} {\bibfnamefont {C.~E.}\ \bibnamefont
  {Doss}}, \bibinfo {author} {\bibfnamefont {E.}~\bibnamefont {Adli}}, \bibinfo
  {author} {\bibfnamefont {R.}~\bibnamefont {Ariniello}}, \bibinfo {author}
  {\bibfnamefont {J.}~\bibnamefont {Cary}}, \bibinfo {author} {\bibfnamefont
  {S.}~\bibnamefont {Corde}}, \bibinfo {author} {\bibfnamefont
  {B.}~\bibnamefont {Hidding}}, \bibinfo {author} {\bibfnamefont {M.~J.}\
  \bibnamefont {Hogan}}, \bibinfo {author} {\bibfnamefont {K.}~\bibnamefont
  {Hunt-Stone}}, \bibinfo {author} {\bibfnamefont {C.}~\bibnamefont {Joshi}},
  \bibinfo {author} {\bibfnamefont {K.~A.}\ \bibnamefont {Marsh}},  \emph
  {et~al.},\ }\bibfield  {title} {\enquote {\bibinfo {title} {Laser-ionized,
  beam-driven, underdense, passive thin plasma lens},}\ }\href {\doibase
  10.1103/PhysRevAccelBeams.22.111001} {\bibfield  {journal} {\bibinfo
  {journal} {Phys. Rev. Accel. Beams}\ }\textbf {\bibinfo {volume} {22}},\
  \bibinfo {pages} {111001} (\bibinfo {year} {2019})}\BibitemShut {NoStop}%
\bibitem [{\citenamefont {Zhu}\ \emph {et~al.}(2023{\natexlab{a}})\citenamefont
  {Zhu}, \citenamefont {Liu}, \citenamefont {Chen}, \citenamefont {Weng},
  \citenamefont {Wu}, \citenamefont {Yu}, \citenamefont {Wang}, \citenamefont
  {Sheng},\ and\ \citenamefont {Zhang}}]{zhu2023magnetic}%
  \BibitemOpen
  \bibfield  {author} {\bibinfo {author} {\bibfnamefont {X.-L.}\ \bibnamefont
  {Zhu}}, \bibinfo {author} {\bibfnamefont {W.-Y.}\ \bibnamefont {Liu}},
  \bibinfo {author} {\bibfnamefont {M.}~\bibnamefont {Chen}}, \bibinfo {author}
  {\bibfnamefont {S.-M.}\ \bibnamefont {Weng}}, \bibinfo {author}
  {\bibfnamefont {D.}~\bibnamefont {Wu}}, \bibinfo {author} {\bibfnamefont
  {T.-P.}\ \bibnamefont {Yu}}, \bibinfo {author} {\bibfnamefont {W.-M.}\
  \bibnamefont {Wang}}, \bibinfo {author} {\bibfnamefont {Z.-M.}\ \bibnamefont
  {Sheng}}, \ and\ \bibinfo {author} {\bibfnamefont {J.}~\bibnamefont
  {Zhang}},\ }\bibfield  {title} {\enquote {\bibinfo {title} {Magnetic pinching
  of relativistic particle beams: a new approach to strong-field {QED}
  physics},}\ }\href {\doibase 10.1088/1367-2630/acf153} {\bibfield  {journal}
  {\bibinfo  {journal} {New J. Phys.}\ }\textbf {\bibinfo {volume} {25}},\
  \bibinfo {pages} {093016} (\bibinfo {year} {2023}{\natexlab{a}})}\BibitemShut
  {NoStop}%
\bibitem [{\citenamefont {Heitler}(1954)}]{heitler1954quantum}%
  \BibitemOpen
  \bibfield  {author} {\bibinfo {author} {\bibfnamefont {W.}~\bibnamefont
  {Heitler}},\ }\href@noop {} {\emph {\bibinfo {title} {The Quantum Theory of
  Radiation}}}\ (\bibinfo  {publisher} {Clarendon Press, Oxford},\ \bibinfo
  {year} {1954})\BibitemShut {NoStop}%
\bibitem [{\citenamefont {Milione}\ \emph {et~al.}(2011)\citenamefont
  {Milione}, \citenamefont {Sztul}, \citenamefont {Nolan},\ and\ \citenamefont
  {Alfano}}]{milione2011higher}%
  \BibitemOpen
  \bibfield  {author} {\bibinfo {author} {\bibfnamefont {G.}~\bibnamefont
  {Milione}}, \bibinfo {author} {\bibfnamefont {H.~I.}\ \bibnamefont {Sztul}},
  \bibinfo {author} {\bibfnamefont {D.~A.}\ \bibnamefont {Nolan}}, \ and\
  \bibinfo {author} {\bibfnamefont {R.~R.}\ \bibnamefont {Alfano}},\ }\bibfield
   {title} {\enquote {\bibinfo {title} {Higher-order {Poincar{\'e}} sphere,
  stokes parameters, and the angular momentum of light},}\ }\href {\doibase
  10.1103/PhysRevLett.107.053601} {\bibfield  {journal} {\bibinfo  {journal}
  {Phys. Rev. Lett.}\ }\textbf {\bibinfo {volume} {107}},\ \bibinfo {pages}
  {053601} (\bibinfo {year} {2011})}\BibitemShut {NoStop}%
\bibitem [{\citenamefont {Sampath}\ \emph {et~al.}(2021)\citenamefont
  {Sampath}, \citenamefont {Davoine}, \citenamefont {Corde}, \citenamefont
  {Gremillet}, \citenamefont {Gilljohann}, \citenamefont {Sangal},
  \citenamefont {Keitel}, \citenamefont {Ariniello}, \citenamefont {Cary},
  \citenamefont {Ekerfelt} \emph {et~al.}}]{sampath2021extremely}%
  \BibitemOpen
  \bibfield  {author} {\bibinfo {author} {\bibfnamefont {A.}~\bibnamefont
  {Sampath}}, \bibinfo {author} {\bibfnamefont {X.}~\bibnamefont {Davoine}},
  \bibinfo {author} {\bibfnamefont {S.}~\bibnamefont {Corde}}, \bibinfo
  {author} {\bibfnamefont {L.}~\bibnamefont {Gremillet}}, \bibinfo {author}
  {\bibfnamefont {M.}~\bibnamefont {Gilljohann}}, \bibinfo {author}
  {\bibfnamefont {M.}~\bibnamefont {Sangal}}, \bibinfo {author} {\bibfnamefont
  {C.~H.}\ \bibnamefont {Keitel}}, \bibinfo {author} {\bibfnamefont
  {R.}~\bibnamefont {Ariniello}}, \bibinfo {author} {\bibfnamefont
  {J.}~\bibnamefont {Cary}}, \bibinfo {author} {\bibfnamefont {H.}~\bibnamefont
  {Ekerfelt}},  \emph {et~al.},\ }\bibfield  {title} {\enquote {\bibinfo
  {title} {Extremely dense gamma-ray pulses in electron beam-multifoil
  collisions},}\ }\href {\doibase 10.1103/PhysRevLett.126.064801} {\bibfield
  {journal} {\bibinfo  {journal} {Phys. Rev. Lett.}\ }\textbf {\bibinfo
  {volume} {126}},\ \bibinfo {pages} {064801} (\bibinfo {year}
  {2021})}\BibitemShut {NoStop}%
\bibitem [{\citenamefont {Zhu}\ \emph {et~al.}(2023{\natexlab{b}})\citenamefont
  {Zhu}, \citenamefont {Liu}, \citenamefont {Chen}, \citenamefont {Weng},
  \citenamefont {Wu}, \citenamefont {Sheng},\ and\ \citenamefont
  {Zhang}}]{zhu2023efficient}%
  \BibitemOpen
  \bibfield  {author} {\bibinfo {author} {\bibfnamefont {X.-L.}\ \bibnamefont
  {Zhu}}, \bibinfo {author} {\bibfnamefont {W.-Y.}\ \bibnamefont {Liu}},
  \bibinfo {author} {\bibfnamefont {M.}~\bibnamefont {Chen}}, \bibinfo {author}
  {\bibfnamefont {S.-M.}\ \bibnamefont {Weng}}, \bibinfo {author}
  {\bibfnamefont {D.}~\bibnamefont {Wu}}, \bibinfo {author} {\bibfnamefont
  {Z.-M.}\ \bibnamefont {Sheng}}, \ and\ \bibinfo {author} {\bibfnamefont
  {J.}~\bibnamefont {Zhang}},\ }\bibfield  {title} {\enquote {\bibinfo {title}
  {Efficient generation of collimated multi-{GeV} gamma-rays along solid
  surfaces},}\ }\href {\doibase 10.1364/OPTICA.479951} {\bibfield  {journal}
  {\bibinfo  {journal} {Optica}\ }\textbf {\bibinfo {volume} {10}},\ \bibinfo
  {pages} {118} (\bibinfo {year} {2023}{\natexlab{b}})}\BibitemShut {NoStop}%
\bibitem [{\citenamefont {Cui}\ \emph {et~al.}(2025)\citenamefont {Cui},
  \citenamefont {Wei}, \citenamefont {Lv}, \citenamefont {Wan}, \citenamefont
  {Salamin}, \citenamefont {Cao},\ and\ \citenamefont
  {Li}}]{Cui2025Generation}%
  \BibitemOpen
  \bibfield  {author} {\bibinfo {author} {\bibfnamefont {L.-J.}\ \bibnamefont
  {Cui}}, \bibinfo {author} {\bibfnamefont {K.-J.}\ \bibnamefont {Wei}},
  \bibinfo {author} {\bibfnamefont {C.}~\bibnamefont {Lv}}, \bibinfo {author}
  {\bibfnamefont {F.}~\bibnamefont {Wan}}, \bibinfo {author} {\bibfnamefont
  {Y.~I.}\ \bibnamefont {Salamin}}, \bibinfo {author} {\bibfnamefont {L.-F.}\
  \bibnamefont {Cao}}, \ and\ \bibinfo {author} {\bibfnamefont {J.-X.}\
  \bibnamefont {Li}},\ }\bibfield  {title} {\enquote {\bibinfo {title}
  {Generation of ultracollimated polarized attosecond
  $\ensuremath{\gamma}$-rays via beam instabilities},}\ }\href {\doibase
  10.1103/PhysRevA.111.023505} {\bibfield  {journal} {\bibinfo  {journal}
  {Phys. Rev. A}\ }\textbf {\bibinfo {volume} {111}},\ \bibinfo {pages}
  {023505} (\bibinfo {year} {2025})}\BibitemShut {NoStop}%
\bibitem [{\citenamefont {Di~Piazza}\ \emph {et~al.}(2012)\citenamefont
  {Di~Piazza}, \citenamefont {M{\"u}ller}, \citenamefont {Hatsagortsyan},\ and\
  \citenamefont {Keitel}}]{piazza2012extremely}%
  \BibitemOpen
  \bibfield  {author} {\bibinfo {author} {\bibfnamefont {A.}~\bibnamefont
  {Di~Piazza}}, \bibinfo {author} {\bibfnamefont {C.}~\bibnamefont
  {M{\"u}ller}}, \bibinfo {author} {\bibfnamefont {K.~Z.}\ \bibnamefont
  {Hatsagortsyan}}, \ and\ \bibinfo {author} {\bibfnamefont {C.~H.}\
  \bibnamefont {Keitel}},\ }\bibfield  {title} {\enquote {\bibinfo {title}
  {Extremely high-intensity laser interactions with fundamental quantum
  systems},}\ }\href {\doibase 10.1103/RevModPhys.84.1177} {\bibfield
  {journal} {\bibinfo  {journal} {Rev. Mod. Phys.}\ }\textbf {\bibinfo {volume}
  {84}},\ \bibinfo {pages} {1177} (\bibinfo {year} {2012})}\BibitemShut
  {NoStop}%
\bibitem [{\citenamefont {Jackson}(2021)}]{jackson2021classical}%
  \BibitemOpen
  \bibfield  {author} {\bibinfo {author} {\bibfnamefont {J.~D.}\ \bibnamefont
  {Jackson}},\ }\href@noop {} {\emph {\bibinfo {title} {Classical
  electrodynamics}}}\ (\bibinfo  {publisher} {John Wiley \& Sons},\ \bibinfo
  {year} {2021})\BibitemShut {NoStop}%
\bibitem [{\citenamefont {Mackenroth}\ and\ \citenamefont
  {Di~Piazza}(2013)}]{mackenroth2013nonlinear}%
  \BibitemOpen
  \bibfield  {author} {\bibinfo {author} {\bibfnamefont {F.}~\bibnamefont
  {Mackenroth}}\ and\ \bibinfo {author} {\bibfnamefont {A.}~\bibnamefont
  {Di~Piazza}},\ }\bibfield  {title} {\enquote {\bibinfo {title} {Nonlinear
  double {Compton} scattering in the ultrarelativistic quantum regime},}\
  }\href {\doibase 10.1103/PhysRevLett.110.070402} {\bibfield  {journal}
  {\bibinfo  {journal} {Phys. Rev. Lett.}\ }\textbf {\bibinfo {volume} {110}},\
  \bibinfo {pages} {070402} (\bibinfo {year} {2013})}\BibitemShut {NoStop}%
\bibitem [{\citenamefont {Li}\ \emph {et~al.}(2015)\citenamefont {Li},
  \citenamefont {Hatsagortsyan}, \citenamefont {Galow},\ and\ \citenamefont
  {Keitel}}]{li2015attosecond}%
  \BibitemOpen
  \bibfield  {author} {\bibinfo {author} {\bibfnamefont {J.-X.}\ \bibnamefont
  {Li}}, \bibinfo {author} {\bibfnamefont {K.~Z.}\ \bibnamefont
  {Hatsagortsyan}}, \bibinfo {author} {\bibfnamefont {B.~J.}\ \bibnamefont
  {Galow}}, \ and\ \bibinfo {author} {\bibfnamefont {Ch.~H.}\ \bibnamefont
  {Keitel}},\ }\bibfield  {title} {\enquote {\bibinfo {title} {Attosecond
  gamma-ray pulses via nonlinear {Compton} scattering in the
  radiation-dominated regime},}\ }\href {\doibase
  10.1103/PhysRevLett.115.204801} {\bibfield  {journal} {\bibinfo  {journal}
  {Phys. Rev. Lett.}\ }\textbf {\bibinfo {volume} {115}},\ \bibinfo {pages}
  {204801} (\bibinfo {year} {2015})}\BibitemShut {NoStop}%
\bibitem [{\citenamefont {Yu}\ \emph {et~al.}(2024)\citenamefont {Yu},
  \citenamefont {Liu}, \citenamefont {Zhao}, \citenamefont {Zhu}, \citenamefont
  {Lu}, \citenamefont {Cao}, \citenamefont {Zhang}, \citenamefont {Shao},\ and\
  \citenamefont {Sheng}}]{yu2024bright}%
  \BibitemOpen
  \bibfield  {author} {\bibinfo {author} {\bibfnamefont {T.-P.}\ \bibnamefont
  {Yu}}, \bibinfo {author} {\bibfnamefont {K.}~\bibnamefont {Liu}}, \bibinfo
  {author} {\bibfnamefont {J.}~\bibnamefont {Zhao}}, \bibinfo {author}
  {\bibfnamefont {X.-L.}\ \bibnamefont {Zhu}}, \bibinfo {author} {\bibfnamefont
  {Y.}~\bibnamefont {Lu}}, \bibinfo {author} {\bibfnamefont {Y.}~\bibnamefont
  {Cao}}, \bibinfo {author} {\bibfnamefont {H.}~\bibnamefont {Zhang}}, \bibinfo
  {author} {\bibfnamefont {F.-Q.}\ \bibnamefont {Shao}}, \ and\ \bibinfo
  {author} {\bibfnamefont {Z.-M.}\ \bibnamefont {Sheng}},\ }\bibfield  {title}
  {\enquote {\bibinfo {title} {Bright {X}/$\gamma$-ray emission and lepton pair
  production by strong laser fields: {A} review},}\ }\href {\doibase
  10.1007/s41614-024-00158-3} {\bibfield  {journal} {\bibinfo  {journal} {Rev.
  Mod. Plasma Phys.}\ }\textbf {\bibinfo {volume} {8}},\ \bibinfo {pages} {24}
  (\bibinfo {year} {2024})}\BibitemShut {NoStop}%
\bibitem [{\citenamefont {Zhao}\ \emph {et~al.}(2019)\citenamefont {Zhao},
  \citenamefont {Liu}, \citenamefont {Li},\ and\ \citenamefont
  {Xia}}]{zhao2019ultra}%
  \BibitemOpen
  \bibfield  {author} {\bibinfo {author} {\bibfnamefont {Y.}~\bibnamefont
  {Zhao}}, \bibinfo {author} {\bibfnamefont {J.-X.}\ \bibnamefont {Liu}},
  \bibinfo {author} {\bibfnamefont {Y.-M.}\ \bibnamefont {Li}}, \ and\ \bibinfo
  {author} {\bibfnamefont {G.-X.}\ \bibnamefont {Xia}},\ }\bibfield  {title}
  {\enquote {\bibinfo {title} {{Ultra-bright $\gamma$-ray emission by using PW
  laser irradiating solid target obliquely}},}\ }\href {\doibase
  10.1088/1361-6587/ab132e} {\bibfield  {journal} {\bibinfo  {journal} {Plasma
  Phys. Control. Fusion}\ }\textbf {\bibinfo {volume} {61}},\ \bibinfo {pages}
  {065010} (\bibinfo {year} {2019})}\BibitemShut {NoStop}%
\bibitem [{\citenamefont {Schwinger}(1951)}]{schwinger1951gauge}%
  \BibitemOpen
  \bibfield  {author} {\bibinfo {author} {\bibfnamefont {J.}~\bibnamefont
  {Schwinger}},\ }\bibfield  {title} {\enquote {\bibinfo {title} {On gauge
  invariance and vacuum polarization},}\ }\href {\doibase
  10.1103/PhysRev.82.664} {\bibfield  {journal} {\bibinfo  {journal} {Phys.
  Rev.}\ }\textbf {\bibinfo {volume} {82}},\ \bibinfo {pages} {664} (\bibinfo
  {year} {1951})}\BibitemShut {NoStop}%
\bibitem [{\citenamefont {Ridgers}\ \emph {et~al.}(2012)\citenamefont
  {Ridgers}, \citenamefont {Brady}, \citenamefont {Duclous}, \citenamefont
  {Kirk}, \citenamefont {Bennett}, \citenamefont {Arber}, \citenamefont
  {Robinson},\ and\ \citenamefont {Bell}}]{ridgers2012dense}%
  \BibitemOpen
  \bibfield  {author} {\bibinfo {author} {\bibfnamefont {C.~P.}\ \bibnamefont
  {Ridgers}}, \bibinfo {author} {\bibfnamefont {Christopher~S.}\ \bibnamefont
  {Brady}}, \bibinfo {author} {\bibfnamefont {R.}~\bibnamefont {Duclous}},
  \bibinfo {author} {\bibfnamefont {J.~G.}\ \bibnamefont {Kirk}}, \bibinfo
  {author} {\bibfnamefont {K.}~\bibnamefont {Bennett}}, \bibinfo {author}
  {\bibfnamefont {T.~D.}\ \bibnamefont {Arber}}, \bibinfo {author}
  {\bibfnamefont {A.~P.~L.}\ \bibnamefont {Robinson}}, \ and\ \bibinfo {author}
  {\bibfnamefont {A.~R.}\ \bibnamefont {Bell}},\ }\bibfield  {title} {\enquote
  {\bibinfo {title} {Dense electron-positron plasmas and ultraintense
  $\ensuremath{\gamma}$ rays from laser-irradiated solids},}\ }\href {\doibase
  10.1103/PhysRevLett.108.165006} {\bibfield  {journal} {\bibinfo  {journal}
  {Phys. Rev. Lett.}\ }\textbf {\bibinfo {volume} {108}},\ \bibinfo {pages}
  {165006} (\bibinfo {year} {2012})}\BibitemShut {NoStop}%
\bibitem [{\citenamefont {Di~Piazza}(2016)}]{piazza2016nonlinear}%
  \BibitemOpen
  \bibfield  {author} {\bibinfo {author} {\bibfnamefont {A.}~\bibnamefont
  {Di~Piazza}},\ }\bibfield  {title} {\enquote {\bibinfo {title} {Nonlinear
  {Breit-Wheeler} pair production in a tightly focused laser beam},}\ }\href
  {\doibase 10.1103/PhysRevLett.117.213201} {\bibfield  {journal} {\bibinfo
  {journal} {Phys. Rev. Lett.}\ }\textbf {\bibinfo {volume} {117}},\ \bibinfo
  {pages} {213201} (\bibinfo {year} {2016})}\BibitemShut {NoStop}%
\bibitem [{\citenamefont {Xie}\ \emph {et~al.}(2017)\citenamefont {Xie},
  \citenamefont {Li},\ and\ \citenamefont {Tang}}]{xie2017electron}%
  \BibitemOpen
  \bibfield  {author} {\bibinfo {author} {\bibfnamefont {B.-S.}\ \bibnamefont
  {Xie}}, \bibinfo {author} {\bibfnamefont {Z.-L.}\ \bibnamefont {Li}}, \ and\
  \bibinfo {author} {\bibfnamefont {S.}~\bibnamefont {Tang}},\ }\bibfield
  {title} {\enquote {\bibinfo {title} {{Electron-positron pair production in
  ultrastrong laser fields}},}\ }\href {\doibase 10.1016/j.mre.2017.07.002}
  {\bibfield  {journal} {\bibinfo  {journal} {Matter Radiat. Extremes}\
  }\textbf {\bibinfo {volume} {2}},\ \bibinfo {pages} {225} (\bibinfo {year}
  {2017})}\BibitemShut {NoStop}%
\bibitem [{\citenamefont {Vranic}\ \emph {et~al.}(2018)\citenamefont {Vranic},
  \citenamefont {Klimo}, \citenamefont {Korn},\ and\ \citenamefont
  {Weber}}]{vranic2018mult}%
  \BibitemOpen
  \bibfield  {author} {\bibinfo {author} {\bibfnamefont {M.}~\bibnamefont
  {Vranic}}, \bibinfo {author} {\bibfnamefont {O.}~\bibnamefont {Klimo}},
  \bibinfo {author} {\bibfnamefont {G.}~\bibnamefont {Korn}}, \ and\ \bibinfo
  {author} {\bibfnamefont {S.}~\bibnamefont {Weber}},\ }\bibfield  {title}
  {\enquote {\bibinfo {title} {{Multi-GeV electron-positron beam generation
  from laser-electron scattering}},}\ }\href {\doibase
  10.1038/s41598-018-23126-7} {\bibfield  {journal} {\bibinfo  {journal} {Sci.
  Rep.}\ }\textbf {\bibinfo {volume} {8}},\ \bibinfo {pages} {4702} (\bibinfo
  {year} {2018})}\BibitemShut {NoStop}%
\bibitem [{\citenamefont {Zhu}\ \emph {et~al.}(2016)\citenamefont {Zhu},
  \citenamefont {Yu}, \citenamefont {Sheng}, \citenamefont {Yin}, \citenamefont
  {Turcu},\ and\ \citenamefont {Pukhov}}]{zhu2016dense}%
  \BibitemOpen
  \bibfield  {author} {\bibinfo {author} {\bibfnamefont {X.-L.}\ \bibnamefont
  {Zhu}}, \bibinfo {author} {\bibfnamefont {T.-P.}\ \bibnamefont {Yu}},
  \bibinfo {author} {\bibfnamefont {Z.-M.}\ \bibnamefont {Sheng}}, \bibinfo
  {author} {\bibfnamefont {Y.}~\bibnamefont {Yin}}, \bibinfo {author}
  {\bibfnamefont {I.~C.~E.}\ \bibnamefont {Turcu}}, \ and\ \bibinfo {author}
  {\bibfnamefont {A.}~\bibnamefont {Pukhov}},\ }\bibfield  {title} {\enquote
  {\bibinfo {title} {{Dense {GeV} electron–positron pairs generated by lasers
  in near-critical-density plasmas}},}\ }\href {\doibase 10.1038/ncomms13686}
  {\bibfield  {journal} {\bibinfo  {journal} {Nat. Commun.}\ }\textbf {\bibinfo
  {volume} {7}},\ \bibinfo {pages} {13686} (\bibinfo {year}
  {2016})}\BibitemShut {NoStop}%
\end{thebibliography}%

\end{document}